\documentstyle[amscd,amssymb,verbatim,euscript]{amsart}
\theoremstyle{plain}
\newtheorem{Thm}{Theorem}[section]
\newtheorem{Prop}[Thm]{Proposition}
\newtheorem{Main}{Main Theorem}

\newtheorem{Lemma}[Thm]{Lemma}

\newtheorem{Def}[Thm]{Definition}

\newtheorem{Exam}{Examples}

\newcommand{\C}[1]{{\mathcal #1}}
\newcommand{\E}[1]{{\EuScript #1}}
\newcommand{\fra}[1]{{\frak #1}}
\newcommand{\supp}{{\rm supp}}
\newcommand{\WF}{{\rm WF}}

\renewcommand{\Bbb}[1]{{\mathbb #1}}

\newenvironment{properties}[1]{\begin{list}%
{\bf P#1.} {\setlength{\itemindent}{-.9pt}}
}
{\end{list}}

\newenvironment{mproperties}[1]{\begin{list}%
{\bf M#1.} 
}
{\end{list}}
\newcommand{\wick}[2]{{:\!\varphi^{#1}(#2)\!:}}
\newcommand{\prodwick}[4]{{:\!\prod_{#1}^{#2}\varphi^{#3}({#4})\!:}}
\newcommand{\multiwick}[4]{{:\!\varphi^{#1}(#2)\cdots\varphi^{#3}(#4)\!:}}
\newcommand{\udot}[1]{{\dot{#1}}}
\newcommand{\singsupp}{\operatorname{singsupp}}
\newcommand{\Endo}{\operatorname{\tt End}}

\newcommand{\id}{\operatorname{id}}
\newcommand{\sd}{\operatorname{sd}}
\newcommand{\msd}{\operatorname{\mu sd}}

\newcommand{\codim}{\operatorname{codim}}
\newcommand{\timo}{\operatorname{to}}
\newcommand{\meno}[2]{{#1\!\setminus\! #2}}
\begin{document}

\title[Microlocal Analysis and Renormalization]{\large\bf Microlocal Analysis 
and Interacting Quantum Field Theories:\\
Renormalization on Physical Backgrounds}
\author[R. Brunetti {\tiny and} K. Fredenhagen]{{\large Romeo Brunetti} 
{\tiny and} {\large Klaus Fredenhagen}\\
$\phantom{}$\\
$\phantom{}$\\
Institut f\"ur Theoretische Physik\\
Universit\"at Hamburg\\
149 Luruper Chaussee\\
D-22761 Hamburg Germany\\
$\phantom{}$\\
{\tt brunetti@@x4u2.desy.de}, {\tt fredenha@@x4u2.desy.de}}

\dedicatory{Dedicated to the memory of Professor Roberto Stroffolini}

\begin{abstract}
We present a perturbative construction of interacting
quantum field theories on smooth globally hyperbolic (curved)
space-times. 
We develop a purely local version of the 
St\"uckelberg-Bogoliubov-Epstein-Glaser method of 
renormalization by using techniques from 
microlocal analysis.
Relying on recent results of Radzikowski,  
K\"ohler and the authors about a formulation 
of a local spectrum condition in terms of wave
front sets of correlation functions of quantum fields on curved 
space-times, we construct 
time-ordered operator-valued products of Wick 
polynomials of free fields. They serve as  building blocks 
for a local (perturbative) definition of interacting fields. 
Renormalization in this framework amounts to extensions of 
expectation values of time-ordered products to all points of space-time.
The extensions are classified according to a microlocal generalization 
of Steinmann scaling degree corresponding to the degree of divergence in 
other renormalization schemes.
As a result, we prove that the usual perturbative classification 
of interacting quantum field theories holds also on curved 
space-times. Finite renormalizations are deferred to a 
subsequent paper.

As byproducts, we describe a perturbative construction of local algebras
of observables, present a new definition of 
Wick polynomials as opera\-tor-va\-lued distributions on a 
natural domain, and we find a general method for the 
extension of distributions which were
defined on the complement of some surface.   
\end{abstract}

\maketitle

\tableofcontents

\section{Introduction}
The quest of the 
existence of a non trivial quantum field theory in four space-time 
dimensions is still without any conclusive result.
Nonetheless, physicists are working daily, with success, on concrete models 
which describe very efficiently physics at wide energy scales.
This description is based on  expansions  of physical
quantities like amplitudes of a scattering process in terms of power
series of ``physical'' parameters, as coupling constants, masses, 
charges. The higher order terms of these power series are usually 
ill defined, in a naive approach, but physicists have
learned soon how to make sense out of them through the procedure now
known as {\it renormalization} \cite{B:schwingerbook}. The rigorous 
extension of this procedure to {\it curved Lorentzian} 
space-times will be the main topic of this paper. The question whether 
the power series approximate the corresponding quantities in a full 
quantum field theory goes beyond the scope of this paper and will not 
be touched.   

The renormalization 
procedure on Minkowski space-time led to impressive results in the case 
of quantum electrodynamics \cite{B:iz,B:weinberg}, 
where observable quantities were calculated and agree
with high precision with the experimental values \cite{B:reviewqed}.
Based on this example, a general method of renormalization of
interacting fields was found and successfully applied to the 
standard models of elementary particles.

There is another approach to quantum field theory (``axiomatic quantum 
field theory'') which
assumes the existence of a class of models satisfying 
certain first principles. Under this assumption several structural 
properties could be derived which therefore hold for every model in 
this class. To name some, CPT and Spin-Statistics 
Theorems are among the main successes of this line of thought, 
and nowadays the application of these methods to specific kind of theories, 
like conformally covariant theories on low dimensional 
spaces, is expected to give new insights.
Nevertheless, all these schemes (see for instance \cite{B:buchholz} 
for a recent survey), either analytic \cite{B:sw} or algebraic 
\cite{B:haag}, by now seem to have missed the challenge for the concreteness 
needed by, say, particle physicists. 

A notable exception is the rigorous formulation of perturbation 
theory \cite{B:hepp,B:eg,B:steinmann,B:zimmermann} which may be 
considered as interpolating between the world of 
phenomenological physics and the
mathematical schemes mentioned above.
This point of view has been pioneered, in 
particular, by K.~Hepp \cite{B:hepp}, who gave solid foundations to 
perturbation theory for quantum field theories on Minkowski space-time. 
This philosophy proved to be
correct for instance in constructive quantum
field theory \cite{B:gj} where rigorous renormalization ideas were used 
as fundamental inputs. 

One of the aims of the authors is to put forward 
a formulation of perturbation theory which 
satisfies the needs of axiomatic field theory, 
much in the sense of \cite{B:steinmann}, and is 
at the same time applicable to phenomenology. In 
distinction to earlier approaches we give a 
purely local formulation which is meaningful 
also on curved space-times.   

Our principal result is:
\begin{Main}
All polynomially interacting (scalar) quantum field theories on 
smooth curved globally hyperbolic space-times of dimension four follow 
the same perturbative classification as on Minkowski space-time.
\end{Main}

Before starting the description of our claim we
continue the description of the interplay between perturbation theory 
and rigorous methods.

One of the most puzzling things in physics is that all the 
attempts to include
gravity in the renormalization program failed: More
recent proposals look for theories of a different kind like
string theory \cite{B:gsw} or its generalizations which are hoped to 
describe all known forces, or Ashtekar program \cite{B:ashtekar}.

Because of the large difference between the Planck scale 
($\approx 10^{-33}$ cm), where ``quantum gravity'' effects are expected 
to become important, 
and the scale relevant for the Standard Model \cite{B:smbook} 
($\approx 10^{-17}$ cm), a reasonable approximation should be to consider 
gravity as a {\it classical} background field and therefore 
investigate quantum
field theory in curved space-time. This Ansatz already led to interesting 
results, the most famous being the Hawking radiation of black-holes 
\cite{B:hawking}. A look through the literature (see, e.g., 
\cite{B:wald,B:birrel,B:fulling}), however, shows that 
predominantly {\it free} field theories were treated on curved physical
backgrounds. In fact, most of the papers on interacting quantum 
field theories on curved space-times deal with the (locally) Euclidean 
case and discuss renormalization only for particular Feynman diagrams. 
We are aware of only one 
attempt to a complete proof of renormalizability, that given 
by Bunch \cite{B:bunch} for the case of 
$\lambda\varphi^4$-model. 
His attempt was however confined to the rather special case of 
real analytic 
space-times which allow an analytic continuation
to the (locally) Euclidean situation. It is interesting to 
note that the main technical tool of his paper is a kind of 
local Fourier transformation and that some of his mathematical 
claims can be justified in the
framework of Hadamard ``parametrices,'' both of which belong to 
the powerful techniques of microlocal analysis that we use 
in this paper.

The situation is then uncertain for general smooth space-times 
with the Lore\-nt\-zi\-an signature for the metric. Here, 
to our knowledge, more or less nothing has been done.

Why is the problem of renormalization so difficult on curved space-times?

Again the precise perspective gained from the rigorous approach is helpful.
The main problem is the absence of translation invariance which in the
rigorous schemes plays a decisive r\^{o}le. In general, no global (time-like)
Killing vector field exists (no energy-momentum operator), so there is no 
canonical notion of a vacuum state, which is a central object in most 
formulations of quantum field theory; the spectrum condition (positivity of 
the energy-momentum operator) can not be formulated. There is no general 
connection between quantum field theories on Riemannian and Lorentzian  
space-times corresponding to the Osterwalder-Schrader 
Theorem \cite{B:os}, and the meaning 
of path integrals for quantum field theories on curved space-times is unclear. 
As a result, the rigorous frameworks described before cannot simply be 
generalized, and the more formal description based on Euclidean ideas and path 
integrals does not help much.

On the other hand, physically motivated by Einstein Equivalence Principle
\cite{B:waldgr}, a quick
look at the possible ultraviolet (short distance) divergences indicates that
they are of the same nature as on Minkowski space, so no real obstruction
for renormalization on curved space-times is visible. Despite of the interest
in its own right, renormalization on curved space-time might also trigger
a conceptual revisitation of renormalization theory on Minkowski space
in the light of the principle of locality \cite{B:df}. 

To develop perturbation theory in a form which is suitable to extensions
to curved Lorentzian space-times, we mainly rely on a construction given by
Epstein and Glaser \cite{B:eg} at the beginning of the seventies and on
some improvements suggested later by Stora 
\cite{B:stora}. This construction makes precise 
older ideas of St\"u\-ckel\-berg \cite{B:stuck} and Bogoliubov \cite{B:bos}.
In spite of its elegance it was widely ignored 
(compare, e.g., its neglection in several books on 
quantum field theory, with the exception of \cite{B:iz}). 
Recently, it was further developed and applied
to gauge field theories by Scharf and his collaborators 
\cite{B:scharf}~\footnote{Note that the term ``finite'' in 
Scharf's book refers to the fact that in the 
Epstein and Glaser approach (as in the similar BPHZ 
method) no regularization is necessary. It does 
not mean that the indeterminacy connected with 
the divergence of naive perturbation theory 
disappears.} (after earlier work by \cite{B:bs}).
 
We offer some intuitive 
explanations of the ideas
behind the approach of St\"u\-ckel\-berg-Bogoliubov-Epstein-Glaser 
(we refrain from using an acronym for this and simply call it the 
Epstein and Glaser approach). 
For simplicity we discuss the case of flat Minkowskian space-time.

The basic idea is that, in the asymptotic past and 
future, the interacting quantum fields approach, in a sense 
to be specified, {\it free} fields, i.e. fields 
satisfying linear hyperbolic equations of motion.
For free fields there exists a precise 
construction which can be used for a 
perturbative description of interacting fields.
Now, in a translationally invariant theory, the 
interacting fields approach the asymptotic free 
fields only in a rather weak sense (LSZ-asymptotic 
condition \cite{B:hepp}). 
Moreover, Haag Theorem \cite{B:haag} forbids the construction of 
interacting fields in the vacuum Hilbert space of 
the time-0 free fields. 
In the Epstein and Glaser scheme, these problems 
are, in a first step, circumvented by choosing 
interactions which take place 
only in a bounded region of space-time. Then the scattering 
operator can be defined in the interaction 
picture as the time evolution operator from the 
past, before the interaction was switched on, to
the future, after the interaction was switched off.

A localized interaction here is thought to be a smooth 
function of time, $t$, with compact support with 
values in the local operators associated with the 
free field. In the simplest case it is
$H_{{\rm int}}(t)=\varphi(f_{t})$ where $\varphi$ is a free field, i.e.
an operator-valued distribution on a Hilbert space, and where 
$f_{t}(x)=\delta(x^0-t)f(x)$ for some test 
function $f$. The $S$-matrix is then an 
operator-valued functional $S(f)$ on the test function 
space. The functional equation for the 
evolution operator implies a factorization 
property for the $S$-matrix if the support of the 
interaction (as a function of time) 
consists of disjoint intervals. In the case above 
with the interaction being a linear function of 
the free field we even find the factorization, 
    \begin{equation}
        S(f_{1}+f_{2})=S(f_{1})S(f_{2})\ ,
        \label{a}
    \end{equation}
whenever there exists some $t\in\Bbb R$ such 
that $\supp(f_{1})\subset\{x|x^0\ge t\}$ and 
$\supp(f_{2})\subset\{x|x^0\le t\}$, and where 
$f_{1}$ and $f_{2}$ need not be smooth at
the hypersurface $x^{0}=t$. This stronger factorization 
property is not expected to hold for more singular 
interactions. Instead we require the following 
consequence of (\ref{a}),
   \begin{equation}
        S(f_{1}+f_{2}+f_{3})=S(f_{1}+f_{2})S(f_{2})^{-1}S(f_{2}+f_{3})\ ,
        \label{b}
   \end{equation}
to hold for test functions $f_{1},f_{2},f_{3}$, 
whenever the supports of $f_{1}$ and $f_{3}$ can 
be separated by a Cauchy surface such that 
$\supp(f_{1})$ lies in the future and 
$\supp(f_{3})$ in the past of this surface. 
Together with the normalization condition 
$S(0)={\boldsymbol 1}$ (identity operator on Hilbert space) 
it implies the first mentioned weaker factorization 
condition in the case $f_{2}=0$.       

The functional equation (\ref{b}) has an 
interesting property; if $S$ is a functional solving 
it we get other
solutions $S_{f}$ by defining $S_{f}(g)=S(f)^{-1}S(f+g)$ (the 
{\it relative} $S$-matrices) where $f$ is an arbitrary test function. 
In particular we get commutativity in case
$\supp (g_1)$ and $\supp (g_2)$ are space-like separated,
    \begin{equation}\label{c:locality}
       S_{f}(g_1 + g_2 )= S_{f}(g_2) S_{f}(g_1) = 
       S_{f}(g_1) S_{f}(g_2)\ .
    \end{equation}
Thus the relative $S$-matrices satisfy the locality 
condition required for local observables. 
They serve as generating functionals for the 
interacting fields.

Unfortunately, a construction of $S(f)$ in four dimensions is known 
only in the case of interaction Hamiltonians which 
are linear or quadratic in the free field, but in two dimensions 
Wrezinski \cite{B:wrezinski} proved that, at least in the particular case
of factorizable $f$, $f(t,x)=g(t)h(x)$, such a construction 
is possible for $\varphi^4$.
One therefore mainly relies on the ``infinitesimal'' 
description of the local $S$-matrix $S(g)$ by studying its 
formal power series \cite{B:bourbaki}
expansion in terms of the ``coupling constant'' $g$. 
The connection with the usual formulation may be done via the 
{\it adiabatic limit},
i.e. the limit for $S(g)$ when $g\to 1$ all over space-time 
is the $S$-matrix, or  in  cases where the limit for $S(g)$ does 
not exist due to infrared divergences,
the limit for the vacuum expectation values of
$S_{g}(f)$, $g\to 1$, is the generating functional 
for the time-ordered correlation functions.

The description given so far emphasizes the fact that the 
Epstein and Glaser method is local in spirit, and it might 
be a favorite candidate 
for developing a renormalization theory on curved space-times. 
A closer inspection, however, shows that 
also in this method translation invariance plays an important 
r\^{o}le, both conceptually and technically, and it will require a 
lot of work to replace it by other structures. 
A similar problem was studied by Dosch and M\"uller \cite{B:dm}. 
These authors developed the
Epstein and Glaser method on Minkowski space for 
quantum electrodynamics with
external time independent electromagnetic fields. Their use of the 
Hadamard parametrices for the 
Dirac operator is already much in the spirit of a 
local formulation of
perturbation theory; by the assumption of time 
independence of the external
fields, however, time translation invariance still plays a crucial 
r\^{o}le in their approach. 
As a matter of fact, it will turn out that 
{\it microlocal analysis} 
\cite{B:micro} is ideally suited to carry 
through the program where in particular the concept of the wave 
front set proves to be extremely useful. We note, {\it en passant}, that 
other reseachers \cite{B:dimock,B:iagolnitzerb,B:iagolnitzerp} had previously 
used these tools in quantum field theory and that more recently 
Verch \cite{B:verch2} has developped
a generalization of the concept of wave front sets which 
can be applied in algebraic quantum field theory.

This paper is an extended version of a previous one 
\cite{B:bfroma} 
where we sketched the main ideas. Here on we give all the 
necessary mathematical details. 

The paper is organized as follows: After this introduction, Section 2
provides some useful grounding concepts and fixes the notations. 
Moreover, we present a new construction of Wick polynomials which may
be of independent interest.
In Section 3 we state the
first principles by which we build up the perturbative method on 
smooth curved
globally hyperbolic space-times. The most important change w.r.t. 
the Epstein
and Glaser method is a characterization of the 
singularity structure of the time-ordered numerical distributions
replacing 
translation covariance. In the course of this 
part we show a local version of the so called 
``Theorem 0''
of Epstein and Glaser which provides the necessary 
mathematical properties of 
the building blocks of the construction. In Section 4 we start the
inductive procedure which aims at constructing the 
time-ordered functions up
to the small diagonal of the product manifold $M^n$, where all
``dangerous'' singularities are located. Sections 5 and 6,
have a more mathematical flavour; we introduce the 
concept of scaling
degree at a point, following 
essentially Steinmann \cite{B:steinmann}, 
and its generalizations in terms of microlocal analysis. The main
aim of this section is the description of the extension to all space 
of distributions 
defined on the complement of a submanifold.
These tools are needed 
for the classification and implementation of renormalizability. 
The next, Section 7, contains the end of the 
inductive procedure by which we 
prove the theories with polynomial interactions to follow the 
same perturbative 
classification as on Minkowski space-time.
 
We emphasize that the method of defining the local 
$S$-matrix joins perturbation theory with the more 
abstract algebraic formulation of quantum field 
theory. In fact, we are able to define a unique family 
(net, precosheaf etc.) of $\ast$-algebras of
observables on globally hyperbolic space-times via the idea of the 
local relative $S$-matrices. Section 8
describes this construction which seems to be widely 
unknown, in spite of the fact that it may already be found, in a preliminary
form, in \cite{B:slavnov}. This Section partly justify
the rather abstract starting point of Section 2.

An outlook, Section 9, concludes the paper. Finally, we stress that the
procedure works for general field theories but for simplicity we stick to the 
notationally easiest case of a single scalar (massive) field theory with self
interactions without derivatives.

\section{General Theory of Quantized Fields and Microlocal Analysis}

In order to fix our notations we recall some basic geometrical 
concepts. 
Further details may be found in some books on general relativity and 
Lorentzian geometry
(see, for instance, \cite{B:waldgr} and \cite{B:bee}). 
We shall work on a {\it space-time} $(M,g)$, where by this 
we mean that $M$ is a connected, Hausdorff, boundaryless 
topological space of pure dimension $d\ge 2$ which $(i)$ 
is paracompact, $(ii)$ is equipped with a smooth structure, 
$(iii)$ is endowed with a Lorentzian metric $g_{ab}$, i.e. a smooth 2-cotensor 
of signature $(1,d-1)$, i.e. $(+,-,\cdots,-)$ and $(iv)$ is oriented and 
time oriented. Given the metric we have a canonically associated 
derivative, namely the 
Levi-Civita derivative denoted by $\nabla$ and an associated curvature tensor
$R^{a}_{bcd}$, with $R$ the scalar curvature. The notion of totally 
geodesic submanifold, i.e., that one for which all tangential geodesics 
stay on the submanifold, is used in Section 6.

Some words on notations. 
Sometimes we write a zero section of a vector bundle $\E{B}$ as $\{0\}$ 
some other times to precise that it belongs to that bundle 
we shall write $Z(\E{B})$. However, in order to avoid any abuse, 
we use the notation $\udot{\E{B}}$
to denote the bundle deprived from its zero section, i.e., 
$\meno{\E{B}}{Z(\E{B})}$. 
We shall also use the notation $M^n$, whenever we treat
the $n$-th order cartesian product of a manifold $M$, and by $\Delta_K$, where
$K\subseteq\{1,\dots,n\}$, we mean the 
(smooth, closed) submanifold of $M^n$ for which any of its
points $(p_i,\dots,p_n)$ are such that $p_{k_{1}} = p_{k_{2}}$ for any pair
$k_1 \neq k_2$ in $K$.

The causality principle plays a crucial r\^{o}le in 
our construction. Therefore we restrict our space-times to be 
{\it globally hyperbolic}. This means that $M$ is homeomorphic to 
${\Bbb R}\times\Sigma$
where $\Sigma$ is a $(d-1)$-dimensional topological submanifold of $M$
and for each $t\in{\Bbb R}$, $\{t\}\times\Sigma$ is a (spacelike) 
Cauchy surface. A Cauchy 
surface is a subset of $M$ which every inextendible non spacelike curve 
intersects exactly once. Given a subset $S$
of $M$ we define the causal future/past sets $J^{\pm}(S)$ as the subsets of 
$M$ which consist of all points  $p\in M$  for which there exists some point
$s\in S$ connected to $p$ by a non space-like future/past directed
curve.  If $M$ is globally hyperbolic, the set 
$J^{+}(p)\cap J^{-}(q)$ is compact for any pair $p,q\in M$. Finally, 
if $p\in M$ then 
the induced metric on tangent space $T_p M$ and cotangent space 
$T^{\ast}_p M$ are Minkowskian, and we
define the future/past light-cones $V_{\pm}$ over these spaces 
(based on $p$) in the usual way.

Quantum field theories on more general spaces pose 
consistency problems (see, e.g.,  
Hawking's ``Chronology Protection Conjecture'' 
\cite{B:hawkingcpc} or the divergence of the 
energy momentum tensor at the Cauchy horizon 
observed in \cite{B:krw}). We remark, however, that since our constructions 
will be purely local, one can as well consider a globally hyperbolic
submanifold of {\it any} Lorentzian space-time.

In many concrete cases, exact solutions of the Einstein 
equation, like Minkowski, de Sitter, Schwartzschild  are real 
analytic. In these cases some of our results might be sharpened 
by working with the analytic 
version of microlocal analysis \cite{B:liess}. In this respect, 
we should mention some recent results of Bros, Epstein
and Moschella for a G{\aa}rding-Wightman-like description of quantum field 
theories on de Sitter space-time \cite{B:bem} where analytic function 
techniques play a major r\^{o}le. 

\subsection{Wave front sets and Hadamard states for free fields}
\label{sub:hadamard}
For the (massive) free field $\varphi$ 
satisfying the (generalized) Klein-Gordon equation of motion
\begin{equation}\label{E:wave}
  (\square_{g}+m^2 +\kappa R)\varphi=0\ ,
\end{equation}
where $\square_{g}$ is the d'Alembertian 
(or Laplace-Beltrami) operator w.r.t. the Lorentzian 
metric $g$, $m\ge 0$ and $\kappa\in\Bbb{R}$, 
one may associate an algebra of observables defined in the 
following way: Let $E_{ret}$ resp. $E_{adv}$ be the 
retarded resp. advanced Green functions of the 
Klein-Gordon operator which are {\it uniquely} defined on 
globally hyperbolic space-times, and let 
$E=E_{ret}-E_{adv}$. Then we consider the unital  
$\ast$-algebra ${\frak A}$ which is generated by the 
symbols $\varphi(f)$, $f\in\E{D}(M)$ (space of complex-valued smooth and 
compactly supported functions), with the 
following relations:
\begin{enumerate}
 \item  The map $f\mapsto\varphi(f)$ is linear,
 \item  $\varphi(f)^{\ast}=\varphi(\bar{f})$,
 \item  $[\varphi(f),\varphi(g)]=iE(f\otimes g){\boldsymbol 1}, 
                \qquad\forall f,g\in\E{D}(M)$,
 \item  $\varphi((\square_{g}+m^2 +\kappa R)f)=0, \qquad\forall f\in\E{D}(M)$,
\end{enumerate}
where the symbol $[\varphi(f),\varphi(g)]$ stands for $\varphi(f)\varphi(g)-
\varphi(g)\varphi(f)$ and $\bar{f}$ means complex conjugation.
A state is, by definition, a linear functional $\omega$
on ${\frak A}$ (the {\it expectation value}) which 
is positive (i.e. $\omega(a^{\ast}a)\ge 0$) and 
normalized ($\omega({\boldsymbol 1})=1$). It is 
uniquely determined by a sequence of multilinear 
functionals $\omega_{n}$, $n=0,1,\ldots$ (the {\it 
$n$-point functions}) on the test 
function space $\E{D}(M)$,
\begin{equation}
\omega_{n}(f_{1},\ldots,f_{n})=\omega(\varphi(f_{1})
\cdots\varphi(f_{n}))\ .       
        \label{d}
\end{equation}
We only consider states whose $n$-point functions 
are distributions and restrict furthermore our attention to the states 
called {\it quasi-free}, namely, 
those states whose only non trivial $n$-point 
functions have $n$ even and are generated in terms of the 2-point functions
(see, e.g. \cite{B:haag}). Among them a distinguished 
class is formed by the so called Hadamard 
states (see, e.g. \cite{B:dwb,B:kw}). 
They are thought to be the appropriate analogue
of the concept of the {\it vacuum} which has no 
direct counterpart on generic space times. In fact, they 
are quasifree states whose 2-point functions have a 
prescribed  short-distance behaviour which is 
partially motivated by the fact that it allows the 
definition of the expectation value of the energy-momentum
tensor (see, e.g. \cite{B:wald}). As first 
observed by Radzikowski~\cite{B:radman} the 2-point 
functions of Hadamard states  can be 
characterized in terms of their wave front set. 

To discuss this characterization we need to enter 
into the realm of microlocal 
analysis. We give some motivations to the basic notions of wave front sets 
and present those basic results which are used throughout the paper. 
We leave the reader the 
task to look further into the large existing literature \cite{B:micro}. 
Physicists might, for concreteness, start from the well-written short 
exposition of Junker in \cite{B:junker}, where they can find definitions
and results about pseudodifferential operators, which we hold as known.

We shall denote by $\E{E}(\Bbb{R}^n)$ the space of complex-valued 
smooth functions and by $\E{E}^\prime (\Bbb{R}^n)$ its dual space, 
i.e., the space of compactly supported distributions.

It is a standard result in distribution theory that 
$u\in\E{E}^{\prime}(\Bbb{R}^n)$ is a smooth function iff its Fourier
transform $\widehat{u}$ decays rapidly in Fourier dual space $\Bbb{R}_n$, i.e.
for any integer $N$ there exists a constant $C_N$ such that
$|\widehat{u}(k)|\le C_N (1+|k|)^{-N}$ for all $k\in\udot{\Bbb{R}}_n$, where
$\meno{\Bbb{R}_n}{\{0\}}\doteq\udot{\Bbb{R}}_n$. 
In case $u$ is not smooth the Fourier transform 
may still rapidly decay in certain directions. We 
may describe this set of directions by an open cone 
in $\udot{\Bbb{R}}_n$
and define $\Sigma(u)$ as its complement in $\udot{\Bbb{R}}_n$.  
It is easy to see 
that $\Sigma(\phi u)\subset\Sigma(u)$ when 
$u\in\E{E}^{\prime}(\Bbb{R}^n)$ and
$\phi\in \E{D}(\Bbb{R}^n)$. 
This property suggests a strategy for the 
general case in which $u$ is not of compact support. So, considering 
$u\in\E{D}^{\prime}(\Bbb{R}^n)$ and a point $x\in\Bbb{R}^n$ 
in the support 
of $u$, $\supp(u)\subset\C{O}$, $\C{O}$ open subset of $\Bbb{R}^n$, 
we first
localize $u$ via multiplication with some $\phi\in \E{D}(\Bbb{R}^n)$ 
such that $\phi(x)\neq 0$
and then consider the Fourier transform of $\phi u$, now a distribution of 
compact support.  
We then define the 
set $\Sigma_{x}(u)\doteq\cap_{\phi}\Sigma(\phi u)$ where the 
intersection is 
taken w.r.t. all smooth functions of compact support $\phi$ such that 
$\phi(x)\neq 0$. This may be called the set of {\it singular directions} of
$u$ over $x$. It is empty whenever $x\notin\singsupp(u)$. 

Hence, finally, we
define the ({\it smooth}) {\it wave front set} for 
$u\in\E{D}^{\prime}(\Bbb{R}^n)$ as
$\WF(u)\doteq\{(x,k)\in\Bbb{R}^n\times\udot{\Bbb{R}}_{n}\ |\ 
 k\in\Sigma_{x}(u)\}$. This set is readily seen to be
closed and conic, where the last means that if $k\in\Sigma_{x}(u)$
so do any $\lambda k$ for all $\lambda>0$. 

It is now crucial that the notion of the wave 
front set can be lifted to any smooth manifold $\E{M}$ where 
it is invariantly defined as a subset of the 
cotangent bundle $\udot{T}^{\ast}\E{M}$. 
This covariance under coordinate transformations is what 
gives to the definition its real technical power. 

Among the results which will be important for us we mention that derivatives 
do not enlarge the wave front set of a distribution, i.e. 
$\WF(\partial u)\subseteq\WF(u)$, and the following criterion 
called H\"ormander
criterion for multiplication of distributions:
\begin{mproperties}{1}
\item {\bf Product}. Picking two distributions
$u_1, u_2\in\E{D}^{\prime}(\E{M})$, the
{\it pointwise} product $u_1 u_2$ exists as a {\it bona-fide} 
distribution whenever $\WF(u_1)+\WF(u_2)$ does not intersect the 
zero section $Z(T^{\ast}\E{M})$, i.e., 
if for all covectors $k_i\in\WF(u_i)$,
$i=1,2$, based over the same point one finds that $k_1 + k_2\neq 0$.
Moreover, if $\WF(u_i)\subset\Gamma_i$, $i=1,2$, then 
$\WF(u_1 u_2)\subset \Gamma_1\cup\Gamma_2\cup ( \Gamma_1 + \Gamma_2 )$. 
\end{mproperties}

We shall also refer frequently to a certain continuity property in 
microlocal analysis which in the body of the paper is sometimes called 
``H\"ormander (pseudo) topology.''  It has to 
do with the notion of convergent sequences which respect also wave front
set properties:
\begin{mproperties}{2}
\item {\bf Continuity}. Let 
$\E{D}^{\prime}_{\Gamma}(\E{M})\doteq\{v\in \E{D}^{\prime}(\E{M})\ |\ 
\WF(v)\subset\Gamma\}$, where $\Gamma$ is a closed conic set 
in $\udot{T}^{\ast}\E{M}$. A sequence $\{u_i\}_{i\in\Bbb{N}}\subset
\E{D}^{\prime}_{\Gamma}(\E{M})$ converges to 
$u\in \E{D}^{\prime}_{\Gamma}(\E{M})$  in the 
sense of the H\"ormander (pseudo) topology whenever the 
following two properties hold true:
\item[]\item[(a)] $u_{i}\to u$ weakly$^{\ast}$ 
(i.e. in $\E{D}^{\prime}(\E{M})$),
\item[]\item[(b)] for any properly supported 
pseudodifferential operator $A$ such
that $\mu\supp(A)\cap\Gamma=\emptyset$, 
we have that $Au_i \to Au$ in the sense of $\E{E}(\E{M})$. 
($\mu\supp(A)$ is the projection onto the second component of the wave
front set of the Schwartz kernel of $A$.)
\end{mproperties}
A last property is connected with the sequential continuity, in the sense 
of {\bf M2}, of the operation
of restriction of a distribution to a submanifold:

\begin{mproperties}{3}
\item {\bf Trace}. Let $\E{N}\subset\E{M}$ denote a submanifold, and let $u\in
\E{D}^{\prime}(\E{M})$. Then $u$ can be restricted to the submanifold $\E{N}$
whenever $\WF(u)$ does not intersect the conormal bundle $N^{\ast}\E{N}$ 
of $\E{N}$. Moreover, 
if $\WF(u)\subset\Gamma$, with $\Gamma$ a closed conic set such that 
$\Gamma\cap N^{\ast}\E{N}=\emptyset$, then the operator of restriction
(trace) $\gamma$ can be lifted as a sequentially continuous operator, in the
sense of {\bf M2}, from
$\E{D}^{\prime}_{\Gamma}(\E{M})$ to $\E{D}^{\prime}(\E{N})$.
\end{mproperties}

For later purpose, it is convenient to have a coordinate dependent formulation
of {\bf M2}(b) by using Fourier transforms. Namely, let $x_0\in\E{M}$ and let 
$V$ be an open conical neighbourhood of $\Gamma_{x_0}$, where the last denotes 
a set of covectors associated to the point $x_0$. Choose a chart 
$(\varphi,U)$ at $x_0$ such that $\Gamma_x\subset V$ for all $x\in U$. Let 
$\chi\in \E{D}(U)$ with $\chi(x_0)\ne 0$. Then the Fourier transform of 
$\chi u$, $u\in \E{D}_{\Gamma}'(\E{M})$, is strongly decreasing in the 
complement of $V$, and
\begin{equation}
\sup_{k\notin V}|(\widehat{\chi u_i}-\widehat{\chi u})(k)|(1+|k|)^N 
\rightarrow 0\ ,
\label{E:Hconv}
\end{equation}
for all $N\in\Bbb{N}$ if $u_i\rightarrow u$ in $\E{D}_{\Gamma}'(\E{M})$.
If, on the contrary, the above convergence holds true for all choices of 
$x_0$, $V$, $(\varphi,U)$ and $\chi$, we obtain {\bf M2}(b).  
  
After this disgression into microlocal analysis we 
briefly describe Radzikowski's characterization of 
Hadamard states \cite{B:radman}. 
The idea is to use wave front sets for a formulation of a
spectral condition. The antisymmetric part of the 
2-point function is the commutator function $E$. Its wave 
front set is 
\begin{equation}\label{E:wfcommutator}
\WF(E)=\{(x,k;x^{\prime},-k^{\prime})\in \udot{T}^{\ast} M^{2}
\ |\ (x,k)\sim (x^{\prime},k^{\prime})\}\ .
\end{equation}
Here the equivalence relation $\sim$ means that there exists a null 
geodesic from $x$ to $x^{\prime}$ such that $k$ is coparallel to 
the tangent 
vector of the geodesic and $k^\prime$ is its parallel transport 
from $x_1$ to 
$x_2$. For coinciding points, the relation is defined as consisting of 
the degenerate (i.e., only one point) 
geodesic at $x=x^\prime$ which has covector $k$ still along 
the boundary of the light-cone and $k^\prime\equiv k$. We remark, for 
later purpose, that since only light-like covectors are present, one can
restrict $E$, and, whenever local coordinates are chosen, its
derivative w.r.t. time $\udot{E}$, to any spacelike Cauchy hypersurface.
 
As a result of \cite{B:radman,B:koehler}, the 2-point function of a Hadamard state has a 
wave front set which is just the positive 
frequency part of $\WF(E)$,   
\begin{equation}\label{E:wfhadamscalar}
\WF(\omega_2)=\{(x,k;x^{\prime},-k^{\prime})\in 
\WF(E)\ |\ \ k\in{\overline{V}_{+}}\}\ .
\end{equation}

Since~(\ref{E:wfhadamscalar}) restricts the
singular support of $\omega_2(x_1,x_2)$ to points $x_1$ and $x_2$
which are null related, $\omega_2$ is smooth for all other points.
The smoothness for {\it space-like\/} related points is known to be
true for quantum field theories on Minkowski space satisfying the
spectrum condition by the Bargmann-Hall-Wightman Theorem \cite{B:sw}. For
time-like related points, however, a similar general prediction on the
smoothness does not exist.  

Another deep result from Radzikowski \cite{B:radman} 
shows that the Duistermaat-H\"or\-man\-der \cite{B:dh} distinguished 
parametrices for the Klein-Gordon equation are nothing 
else than the (St\"uckelberg-)Feynman--anti-Feynman propagators 
(up to $C^\infty$) for quasi-free Hadamard states. We recall that the 
time-ordered 2-point function $E_F$ arising from $\omega_2$ is given by
\begin{equation*}
iE_F (x_1,x_2)=\omega_2 (x_1,x_2)+ E_{ret} (x_1,x_2) \ .
\end{equation*}
Its wave front set \cite{B:radman} is
\begin{equation*}
    \WF(E_F)= O\cup D\ ,
\end{equation*}
where the off-diagonal piece is given by,
\begin{equation*}
    O=\{(x,k;x^\prime,-k^\prime)\in\udot{T}^{\ast}M^2\ |\  
    (x,k)\sim (x^\prime,k^\prime), x\neq x^\prime, k\in \overline{V}_{\pm}\ 
    \text{if}\  x\in J^{\pm}(x^\prime)\}\ ,
\end{equation*}
and the diagonal one by, 
\begin{equation*}
    D= \{(x,k;x,-k)\in\udot{T}^{\ast}M^2\ |\  x\in M, k\in 
    \udot{T}^{\ast}_{x} M\}\ .
\end{equation*}

Now, one can see why in naive perturbation theory we may find divergences. 
Indeed, the perturbative expansion in terms of Feynman 
graphs in position space leads to pointwise 
products of Feynman propagators. 
But these products do not satisfy H\"ormander criterion for 
multiplication of distributions  
since covectors based on the diagonal piece $D$ can add up to zero.

\subsection{A new construction of Wick polynomials}\label{sub:wfs}

In a previous paper \cite{B:bfk} 
we constructed Wick polynomials as operator-valued 
distributions.  We considered a fixed Hadamard 
state $\omega$ and the induced GNS representation 
$(\C{H}_{\omega},\pi_{\omega},\Omega_{\omega})$ for the $\ast$-algebra
$\frak A$
and found the Wick polynomials as operator-valued 
distributions on the dense cyclic domain generated 
by $\Omega_{\omega}$. We recall that by a GNS triple
$(\C{H}_{\omega},\pi_{\omega},\Omega_{\omega})$ we mean a complex Hilbert 
space $\C{H}_{\omega}$, a representation $\pi_{\omega}$ of $\frak A$ by
unbounded operators on $\C{H}_{\omega}$, and finally by 
$\Omega_{\omega}\in\C{H}_{\omega}$ the cyclic vector representing the state
$\omega$ for which one has the connection equation
\begin{equation*}
\omega(A)=(\Omega_\omega,\pi_\omega (A) \Omega_\omega )\ , 
\qquad\forall A\in{\frak A}\ .
\end{equation*}

The dependence of the construction of Wick polynomials
on the choice of the Hadamard state led to two 
problems: The first one is due to the convention 
that the expectation value of a Wick polynomial 
vanishes in the chosen Hadamard state. Other 
choices lead to a finite redefinition, a problem 
well known from the definition of the expectation 
value of the energy momentum tensor \cite{B:wald2}. Since we shall not
discuss finite renormalizations in this paper, 
we do not treat this problem at the moment. 
The other problem is of a more technical nature: 
The smeared Wick polynomials are unbounded 
operators. We know from the work of 
Verch \cite{B:verch} that, locally, i.e. in bounded regions of space time, 
different Hadamard states lead to equivalent 
representations. But this theorem does not 
guarantee that the domains of definition for 
different choices of the cyclic vector coincide. 
We therefore give here a new definition which 
depends only on the representation but not on the 
cyclic vector. Its restriction to the cyclic 
subspaces coincides with the previous definition.

It is well known that the operators $\varphi(f)$ (now, 
representatives under $\pi_\omega$ of the abstract elements 
of the $\ast$-algebra $\frak A$ in subSection 2.1) 
for a real valued test function are essentially 
self adjoint on the cyclic domain generated by 
$\Omega_{\omega}$, and that the Weyl operators 
$W(f)=\exp(i\varphi(f)^{\ast\ast})$ (where now the $\ast$-operation 
denotes the Hilbert space adjoint) satisfy the Weyl 
relation $W(f)W(g)=\exp(-\frac{i}{2}E(f\otimes 
g))W(f+g)$. The expectation value in the given 
Hadamard state is
   \begin{equation}
        \omega(W(f))=\exp(-\frac{1}{2}\omega_{2}(f,f))\ .
        \label{E:Weyl}
   \end{equation}
Let $:\!\! W(f)\!\!:\ \doteq 
\exp(\frac{1}{2}\omega_{2}(f,f))W(f)$, and define 
for $\Psi\in \C{H}_{\omega}$ the vector-valued function 
$\Psi(f)=\ :\! W(f)\!:\Psi$ .

\begin{Def}\label{D:fd} 
We say that $\Psi(f)$ is {\rm infinitely often differentiable} 
at $f=0$ if 
there exists for every integer $n\ge 0$ a  
symmetrical vector-valued 
distribution $\delta^n\Psi /\delta f^n$ 
on $\E{D}(M^n)$, 
and continuous seminorms $p_{n}$ on the test function 
space $\E{D}(M)$ with $p_{n+1}\ge p_{n}$, such that
\begin{itemize}
\item[(a)] if $p_{0}(h)=0$ then $\Psi(h)=\Psi(0)\ $,
\item[(b)] if $h\rightarrow 0$ with $p_{n}(h)\ne 0$ then
    \begin{equation*}
        \left\|\Psi(h)-\sum_{l=0}^n\frac{1}{l!}\frac{\delta^l\Psi}{\delta 
        f^l}(h^{\otimes l})\right\|\ p_{n}(h)^{-n}\longrightarrow 0\ , 
        \label{E:diff}
    \end{equation*}
where $\|\ \cdot\ \|$ stands for the Hilbert space norm in $\C{H}_\omega$.
\end{itemize}
\end{Def}

The kernel of the functional derivative can be written,
\begin{equation}
\frac{\delta^n\Psi}{\delta f(x_{1})\cdots\delta 
f(x_{n})}= i^{n} :\!\varphi(x_{1})\cdots\varphi(x_{n})\!:\Psi\ .
        \label{E:wickprod}
\end{equation}
The right hand side
of Eqn.~(\ref{E:wickprod}) defines what is called a Wick monomial.

We want to find those vectors on which the Wick monomials 
can be restricted to partial diagonals. In view of the 
criterion {\bf M3} for the restriction of distributions, we
define as the {\it microlocal 
domain of smoothness} the following set:
\begin{equation}
 \begin{split}
\C{D}=\biggl\{&\Psi\in\C{H}_{\omega}\ |\ \Psi(f)\ \text{is infinitely 
often differentiable at $f=0$, and for every}\\ 
&\text{$n\in\Bbb{N}$ the wavefront set 
of $\frac{\delta^n\Psi}{\delta f^n}$ is contained in the set}\\ 
&\{(x_1, k_1 ;\dots ; x_n,k_n)\in \udot{T}^{\ast} 
M^n\ |\ k_i\in\overline{V}_{-},\ i=1,\dots,n\}\biggr\}\ .
 \end{split}
        \label{E:wickdomain}
\end{equation}    

The vector-valued distributions (\ref{E:wickprod}) with 
$\Psi\in\C{D}$ can be restricted to 
all partial diagonals, and give all possible Wick 
polynomials. Moreover, according to {\bf M1}, they may also 
be
multiplied by distributions whose wavefront sets do not contain 
elements $(x_1,k_1;\dots; x_n,k_n)\in \udot{T}^{\ast}M^n$, 
$k_i\in\overline{V}_{+},\ i=1,\dots,n$.
The domain $\C{D}$ 
is invariant under application of Weyl operators 
and smeared Wick polynomials. 

We state a crucial property:
\begin{Lemma}
Let $\Phi\in\C{H}_\omega$ 
induce some quasi-free Hadamard state $\omega^\prime$. 
Then $\Phi\in\C{D}$.
\end{Lemma}
\begin{pf}
The main point rests on the validity of Leibniz rule. Indeed, we can
write, 
\begin{equation*}
\Phi(f) = \exp(\frac{1}{2}(\omega_2(f,f)-\omega_2^\prime(f,f))
\exp(\frac{1}{2}\omega_2^\prime(f,f)+i\varphi(f)^{\ast\ast})\Phi\ ,
\end{equation*}
and differentiate w.r.t. $f$. The general $n$-th order derivative gives
\begin{equation}\label{E:der}
\frac{\delta^n \Phi}{\delta f^n}(h^{\otimes n}) =
\sum_{I\subseteq\{1,\dots,n\},|I|\mbox{ even}}\chi(h,h)^{|I|/2}\  
\frac{\delta^{|I^c |}\Phi^{\prime}}{\delta f^{| I^c |}}(h^{\otimes |I^c|})\ ,
\end{equation}
where, $\Phi^{\prime} (f)=\exp(\frac{1}{2} \omega_2^{\prime}(f,f)
+i\varphi(f)^{\ast\ast})\Phi$,
$\chi = \omega_2 -\omega_2^\prime$ is a smooth function on $M^2$ and a 
solution of the Klein-Gordon equation in both entries.

Now, $\Phi^{\prime}(h)$ satisfies the estimate in Definition~\ref{D:fd} 
with the seminorms 
$p_{n}^{\prime}(h)=\sqrt{\omega_{2}^{\prime}(h,h)}$ for all $n$, and the 
numerical prefactor with the seminorms 
$p_{n}(h)=\sqrt{(\omega_2 +\omega_2^\prime)(h,h)}$, hence for the 
whole expression we may also use the seminorms $p_{n}$. 
Actually, as was shown by Verch in \cite{B:verch},
there exist two positive constants $A$ and $B$, such that
\begin{equation*}
A\omega_2(f,g)\le\omega_2^\prime (f,g)\le B\omega_2(f,g)\ , \qquad 
f,g\in\E{D}(M)\ ,
\end{equation*}
hence all these seminorms are equivalent. We conclude 
that $\Phi(f)$ is infinitely differentiable at $f=0$.

The wave front sets of the
functional derivatives of $\Phi(f)$ and $\Phi^{\prime}(f)$ 
coincide, since $\chi$ is smooth. Using the formula
\begin{equation}
        \left\|\frac{\delta^n\Phi^{\prime}}{\delta f^n}(h^{\otimes 
        n})\right\|^2=n!\ \omega_{2}^{\prime}(h,h)^n \ ,
        \label{E:wickexpv}
\end{equation}
and the information on the wave front set of Hadamard 
states, we find  
\begin{equation}
        \WF\left(\frac{\delta^n\Phi}{\delta f^n}\right)=
        \{(x_1,k_1; \dots; x_n, k_n)\in \udot{T}^{\ast} 
M^n\ |\ k_i\in \partial V_{-},\ i=1,\dots,n\}\ ,
        \label{E:wickwf}
\end{equation} 
so $\Phi\in\C{D}$.
\end{pf}

We use the so called (generalized) Wick expansion formula which 
is the basic combinatorial formula for perturbation theory. Let us denote
by $\E{L}$ any Wick polynomial on $\C{D}\subset\C{H}_\omega$, and
we define the ``derivative'' of a
Wick polynomial  with respect to $\varphi$ to be $\partial {\E 
L}/\partial\varphi$. We can characterize it by the following result:

\begin{Lemma} There is a unique Wick polynomial $\partial {\E 
L}/\partial\varphi$ which satisfies the equation,
\begin{equation}\label{E:comm}
[{\E L}(x),\varphi(y)]=\frac{\partial {\E 
L}}{\partial\varphi}(x)E(x,y)\ ,
\end{equation} 
in the sense of operator-valued distributions on $\C{D}$.
\end{Lemma}
\begin{pf}
By linearity it is 
sufficient to prove it for any Wick monomial e.g. $\wick{n}{x}$. 
It is obvious, that $n\wick{n-1}{x}$ satisfies (\ref{E:comm}), 
hence we only
need to proof that if $A$ is any Wick polynomial for which $A(x)E(x,y)=0$ 
this means that $A\equiv 0$. But this follows from the fact 
that for every $x\in M$ we can find some 
test function $f$ such that the (smooth) solution $E(x,f)$ of the 
Klein-Gordon equation does not vanish at $x$.
\end{pf}

Now, let us consider, for any Wick polynomial $\E{L}$, the fields
${\E L}^{(j)}= \partial^j {\E L}/\partial\varphi^j$, $j\in\Bbb{N}$, 
which, by induction, are uniquely defined according to the previous Lemma.

\begin{Thm} [Generalized Wick expansion Theorem]\label{T:wick}
Let $\E{L}_k$, $k=1,\dots,n$, be Wick polynomials. The 
following relation holds:
\begin{equation}\label{E:wickex} 
\begin{split} 
{\E L}_{1}(x_1)\cdots{\E L}_{n}(x_n)=\sum_{j_1,\dots,j_n}
(\Omega_\omega, {\E L}_{1}^{(j_{1})}(x_1)\cdots {\E 
L}_{n}^{(j_{n})}&(x_n)
\Omega_\omega)\times\\ 
&\times \frac{\multiwick{j_{1}}{x_1}{j_{n}}{x_n}}{j_{1}!\cdots j_{n}!}\ ,
\end{split}
\end{equation}
where the summations over the $j_{k}$'s go from 0 to the order of the 
corresponding ${\E L}_k$.
\end{Thm}

For a proof see \cite{B:hepp} or just use the previous notion of 
differentiability and apply induction. Note that the products in the 
Theorem above exist because the 
wave front sets of their expectation values satisfy 
H\"ormander criterion {\bf M1} due to the convexity 
of the forward light cone. 

The wave front sets for the Wightman distributions of Wick polynomials
may be larger than those of the 
corresponding distributions for the free field $\varphi$. Consider as
an example the 2-point Wightman function for the Wick monomial 
$\wick{2}{x}$, i.e.
$(\Omega_\omega,\wick{2}{x_1}\wick{2}{x_2}\Omega_\omega)$. According to the 
Theorem, this is equal to $2\omega_{2}(x_1,x_2)^{2}$.  This product 
exists according to the H\"ormander criterion,
and its wave front set is contained in 
$(\WF(\omega_2)+\WF(\omega_2))\cup\WF(\omega_2)\doteq\digamma_2$. The set
$\digamma_2$ will be instrumental for some results below. 
Now, $\WF(\omega_2)+\WF(\omega_2)$ contains directions which lie 
{\it inside} the light-cone as it is clear by adding up two covectors 
$k_1 +k_2$ for points on the diagonal. One thus sees how already 
the smallest possible non-linearity may give rise to additional singular 
directions w.r.t. those already present in the wave front sets of the 
Wightman functions for the original field $\varphi$. Another important
remark is that $\digamma_2$ is an involutive closed cone, i.e. is a closed
cone which is stable under sums, and, as a straightforward result, 
it gives that $(\omega_{2}(x,y))^n$
still has $\WF(\omega_2^n)\subset \digamma_2$.

The general structure for multi-point expectation values of Wick 
polynomials can be found as follows. For a more compact notation
some definitions from graph theory are used: Let $\C{G}_n$
denote the set of all finite nonoriented graphs with vertices 
$V=\{1,\ldots,n\}$ and let $E^G$ denote the set of      
edges of a given graph $G$. Moreover, 
for any vertex $i\in V$ we denote by $E_{i}^{G}$ the subset of edges which 
belong to the vertex $i$, possibly empty, and by $|E_{i}^{G}|$ their number and
similarly by $E_{ij}^{G}$ the subset of edges connecting points $i$ and $j$, 
with the obvious relation $|E^{G}_{i}|=\sum_{j} |E_{ij}^{G}|$.
For any edge $e\in E$ connecting points $i$ and $j$ 
we use the ``source and range'' notation, i.e. $i=s(e)$ and $j=r(e)$, 
whenever $i<j$. 

It is sufficient, by linearity, to restrict ourselves to the treatment
of products of Wick monomials. Indeed, let us denote by 
$\omega_{n}^{m_1,\dots, m_n}$ the expectational value, w.r.t. the GNS-vector
$\Omega_\omega$ for a quasi-free Hadamard state $\omega$, of the
product of Wick monomials $\wick{m_1}{x_1}\cdots\wick{m_n}{x_n}\/$ ,
and define as $\widehat{\C{G}}_{n}(m_1,\dots,m_n)$ the set of all graphs
$G$ for which all vertices $j$ with $m_j$ edges are saturated, 
i.e. $|E^{G}_j|=m_j$. Moreover, following \cite{B:bfk}, 
we call a triple $(x,\gamma,k)$ an 
{\it immersion} of any graph $G\subset\C{G}_n$ into the manifold $M$ whenever, 
($a$) $x: V\rightarrow M$ is a map from all vertices $i$ of $G$ to points $x_i$
of $M$; ($b$) $\gamma$ maps edges $e\in E^G$ to null geodesics $\gamma_e$
connecting points $x_{s(e)}$ and $x_{r(e)}$; ($c$) $k$ maps edges 
$e\in E^G$ to future 
directed covector fields $k_{\gamma_{(e)}}\equiv k_e$ 
which are coparallel to the
tangent vector $\dot{\gamma}_{e}$ of the null geodesic.
Hence, generalizing the set $\digamma_2$ above by,
\begin{equation}
  \begin{split}
    \digamma_{n}\doteq \biggl\{&(x_1,k_1;\ldots;x_n,k_n)\in
      \udot{T}^{\ast} M^n| \quad 
        \exists\ G\in\widehat{\C{G}}_n(m_1,\dots,m_n)\ \text{and an}\\
  & \qquad\text{immersion $(x,\gamma,k)$ of $G$ such that}\\
  & \qquad\qquad    k_i =\sum_{\begin{Sb}
                           e\in E^{G}\\
                           s(e)=i 
                         \end{Sb}
                        } k_e (x_i)
                   -\sum_{\begin{Sb}
                           e\in E^{G}\\
                           r(e)=i 
                         \end{Sb}
                        } k_e (x_i)
  \biggr\}\ ,\\
  \end{split}
\end{equation}
we have:
\begin{Prop}
The wave front set of $\omega_n^{m_1,\dots,m_n}$ is geometrically bounded
as
\begin{equation}
\WF(\omega_n^{m_1,\dots,m_n})\subset\digamma_n\ .
\end{equation}
\end{Prop}
\begin{pf}
Let us define for notational purpose $\widehat{\C{G}}_{n}\doteq
\widehat{\C{G}}_{n}(m_1,\dots,m_n)$.
As follows from Theorem~\ref{T:wick} the expectation value is,
\begin{equation}
\omega_n^{m_1,\ldots,m_n}(x_1,\dots,x_n)\doteq
\sum_{G\in\widehat{\C{G}}_n} \omega_n^{G}(x_1,\dots,x_n)=
\sum_{G\in\widehat{\C{G}}_n} \prod_{e\in G}\omega_{2}(x_{s(e)},x_{r(e)})\ .
\end{equation}
Considering one graph $G$ in this sum we see that, in explicit form,
\begin{equation*}
\begin{split}
\omega_n^{G}(x_1,\dots,x_n)&\doteq\prod_{e\in G}\omega_{2}(x_{s(e)},x_{r(e)})\\
&=\omega_2(x_1,x_2)^{|E_{1,2}^{G}|}\omega_2(x_1,x_3)^{|E_{1,3}^{G}|}\cdots
\omega_2(x_{n-1},x_{n})^{|E_{n-1,n}^{G}|}.
\end{split}
\end{equation*}
In the last equality the 2-point distributions should be understood as 
distributions on the product manifold $M^n$. Hence their wave front set
is given, when $i <j$, by,
\begin{equation*}
\WF(\omega_2^{|E_{i,j}^{G}|})\subset
\{(x_1,0;\dots;x_i,k_{i,j};\dots; x_j,-k_{i,j};\dots;x_n,0)| 
(x_i,k_{i,j};x_j,-k_{i,j})\in\digamma_2\}.
\end{equation*}

It is straightforward to see from the last expression that the claim of 
the Proposition is correct. Indeed a general covector $k_i$ will have the 
following expression,
\begin{equation}\label{E:sumofcov}
k_i = -k_{1,i}-\cdots - k_{i-1,i} + k_{i, i+1} + \dots + k_{i,n}\ ,
\end{equation}
where some of the $k_{l,m}$ may be zero.
Now, as follows from Eqn.~(\ref{E:wfhadamscalar}), 
to any edge $e$ which consists of pair of joined vertices $i, j$ 
in the graph $G$ there exist points on the manifold $x_i, x_j$ and 
a null geodesic $\gamma_e$ connecting them together with 
a future directed covector
field $k_e$ which is coparallel to the tangent vector of the geodesic 
and is such that, in agreement with Eqn.~(\ref{E:sumofcov}), 
\begin{equation*}
k_i =\sum_{\begin{Sb}e\in E^{G}\\s(e)=i\end{Sb}} k_e (x_i)
                   -\sum_{\begin{Sb}e\in E^{G}\\
                           r(e)=i 
                         \end{Sb}
                        } k_e (x_i)\ .
\end{equation*}
Since this applies equally well to all graphs, and since the wave front set
of sums of distributions is bounded by the union of the wave front sets, 
we get the thesis.
\end{pf}

\section{On a Local Formulation of Perturbation Theory}

We have recalled in the introduction the main 
ideas of the Epstein and Glaser formulation of perturbation theory. 
Here we give the details of our generalization. 

We start from the Gell-Mann and Low formula for the $S$-matrix for 
quantum theories on four dimensional Minkowski space-time $\Bbb{M}$; 
this means adopting the following formal expression,
\begin{equation*}
S_{\lambda} = T(e^{i\lambda\int_{\Bbb M} {\E L}_{int}(x) d^4 x})\ ,
\end{equation*}
where $T$ denotes the notion of time ordering, ${\E L}_{int}$, the 
interaction Lagrangian, is some local field   and $\lambda$ is the
strength of the interaction. Developing in Taylor series
w.r.t. $\lambda$  gives
\begin{equation*}
S_{\lambda} = \sum_{k=0}^{\infty}\frac{(i\lambda)^k}{k!} 
\int_{\Bbb M}\cdots\int_{\Bbb M} T({\E L}_{int}(x_1){\E L}_{int}(x_2)\cdots
{\E L}_{int}(x_k)) \prod_{i=1}^{k} d^4 x_i\ .
\end{equation*}
Hence, the perturbative solution to scattering theory is reduced to 
quadratures once one finds 
the  general solution of the time ordering operation 
inside the integral. For noncoinciding points the 
solution is given by the following expression,
\begin{equation}
\begin{split}
T({\E L}_{int}&(x_1)\cdots
{\E L}_{int}(x_n))=\\
&\sum_{\pi\in{\E P}_{1,\dots,n}} \theta(x_{\pi(1)}-x_{\pi(2)})
\cdots\theta(x_{\pi(n-1)}-x_{\pi(n)}) {\E L}_{int}(x_{\pi(1)})\cdots
{\E L}_{int}(x_{\pi(n)})\ ,
\end{split}
\label{E:naivetimeordering}
\end{equation}
where ${\E P}_{1,\dots,n}$ is the set of all permutations of the index set 
$\{1, \dots, n\}$ and $\theta$ is the Heaviside step function,
\begin{equation*}
\theta(x)=\begin{cases} 1, &\text{if}\ x^{0}>0,\\
                        0, &\text{otherwise},
          \end{cases}
\end{equation*}
where $x^0$ denotes the time component of the points 
in $\Bbb M$. As well known, 
this expression leads to the description of scattering processes 
by Feynman graphs \cite{B:dyson}. Due to local 
commutativity of the Lagrangian the singularities 
of the Heaviside step function at coinciding times 
are harmless, as long as all points are different.
Unfortunately, this is no longer true for 
coincident points, since ${\E L}_{int}$ is an operator-valued 
distribution which cannot, in general, be multiplied 
by a discontinuous function; if one tries to 
define the products by convolutions in momentum 
space, this leads, in a 
naive approach, to the occurrence of ultraviolet divergences. 

Several procedures have been found to cope with these singularities. 
But typically they are nonlocal and are therefore not 
immediately generalizable to the case of 
Lorentzian curved backgrounds.
Better is the situation in Euclidean field theory 
(see e.g. \cite{B:luescher} for a 
generalization of dimensional renormalization to 
the curved case). 

Let us consider, as possible interaction Lagrangians ${\E L}$, 
Wick polynomials of the free field $\varphi$. From 
Eqns.~(\ref{E:naivetimeordering}) and
(\ref{E:wickex}) we find
\begin{equation}\label{E:wicktime}
\begin{split} 
T({\E L}_{1}(x_1)\cdots{\E L}_{}(x_n))=\sum_{j_1,\dots,j_n}
(\Omega_\omega,T({\E L}_{1}^{(j_{1})}&(x_1)\cdots {\E L}_{n}^{(j_{n})}(x_n))
\Omega_\omega)\times\\ 
&\times 
\frac{\multiwick{j_{1}}{x_1}{j_{n}}{x_n}}{j_{1}!\cdots
j_{n}!}\ ,
\end{split}
\end{equation}
where, however, the expectation value in the right-hand side 
is {\it a priori} not 
defined all over $\Bbb{M}^n$, but only  over $\meno{\Bbb{M}^n}{\Sigma}$ where 
$\Sigma$ is the union of all diagonals in $\Bbb{M}^n$. So, the 
main problem 
is to give a mathematical meaning to this formula on all 
points. 

\subsection{Formulation of the local $\boldmath S$-matrix} 
The general starting idea, due to Bogoliubov, is 
to consider the usual Gell-Mann and 
Low formulation of the $S$-matrix but supplemented by the 
hypothesis necessary
to the implementation of the causality principle. In one stroke 
one finds also 
the solution to the problem of the correct treatment of the 
operator-valued distributions. Choosing as the interaction 
Lagrangian  ${\E L}_{int}(x)={\E L}(x)\eta(x)$, a Wick polynomial ${\E L}$
multiplied by a space-time function of compact support 
$\eta\in \E{D}(M)$ (considered as a generalized 
``coupling constant''), we {\it define} the local $S$-matrix
$S_{\lambda}(\eta)$ as a formal power series 
(see, for instance, \cite{B:bourbaki}) in the coupling strength $\lambda$,
\begin{equation}\label{E:defsmatrix}
S_{\lambda}(\eta)\doteq {\boldsymbol 1} + 
\sum_{n=1}^{\infty}\frac{(i\lambda)^n}{n!}
\int_{M^n} T({\E L}_{int}(x_{1})\cdots{\E
  L}_{int}(x_{n}))
d\mu_1\cdots d\mu_n\ ,
\end{equation}
where $d\mu$ is the natural invariant volume measure on the globally hyperbolic
space-time $(M,g)$, and ${\boldsymbol 1}$ is the Hilbert space identity 
operator. One can enlarge the definition e.g. by 
using a more general Lagrangian,
\begin{equation}\label{E:lagrangian}
\lambda\E{L}_{int}=
\sum_{k=1}^{l}\eta_{k}\E{L}_{k}\ ,
\end{equation}
with Wick polynomials $\E{L}_{k}$ and associating 
to each a different ``coupling constant''
$\eta_k\in \E{D}(M)$, 
where the additional ``Lagrangians'' $\E{L}_{k}$ are defined as terms like
currents, external fields etc., including in 
particular all derivatives of the basic interaction 
Lagrangian.  Using this extended Lagrangian
we may replace the time-ordered operator in Eqn.~(\ref{E:defsmatrix}) by,
\begin{equation*}
T({\E L}_{int}(x_{1})\cdots{\E
  L}_{int}(x_{n}))=\sum_{k_1,\dots,k_n} 
T({\E L}_{k_{1}}(x_{1})\cdots{\E 
L}_{k_{n}}(x_{n}))
\eta_{k_1}(x_1)\cdots \eta_{k_n}(x_n)\ ,
\end{equation*}
where the summation over the $k$'s go from $1$ to $l$, 
the number of the terms in the extended Lagrangian.

We remark that, eventually, the test function(s) $\eta$ should be sent 
to a fixed value over all space-time. This procedure, 
known as adiabatic limit, amounts to treat the infrared 
nature of the theory. Some studies 
of this limit in the case of Minkowski space-time 
have been performed by 
Epstein and Glaser themselves \cite{B:eginfrared}. 
It is not clear how to generalize their 
study to curved spaces. It is therefore 
gratifying that all local properties of the theory 
are already obtained via the  construction of the local 
$S$-matrices, and this point of view might also be 
useful in cases (like non abelian gauge theories) where due to 
infrared problems the $S$-matrix in the adiabatic limit does not 
properly exist.

\subsection{Defining properties}
Our main goal is the inductive construction of the time-ordered
products of Wick polynomials,
\begin{equation*}
  T({\E L}_{1}(x_{1})\cdots{\E L}_{n}(x_{n}))\ .
\end{equation*}
Following Epstein and Glaser we require the following properties:

\begin{properties}{1}
\item {\bf Well-posedness}. The symbols 
  $\ T({\E L}_{1}(x_{1})\cdots{\E L}_{n}(x_{n}))$ are 
  well defined ope\-ra\-tor-valued 
  distributions on the GNS-Hilbert space $\C{H}_{\omega}$ i.e. 
  (multilinear, strongly continuous) maps 
  $\E{D}(M^n)\rightarrow{\Endo}(\C{D})$ where $\C{D}\subset\C{H}_{\omega}$
  is the dense subspace (microlocal domain of smoothness) 
  defined in (\ref{E:wickdomain}).
\end{properties}

\begin{properties}{2}
\item {\bf Symmetry}.
  Any time-ordered product $\ T({\E L}_{1}(x_{1})\cdots {\E L}_{n}(x_{n}))$ 
  is symmetric under permutations of indices, i.e. the action of the 
  permutation group $\E{P}_{\{1,\dots,n\}}$ of the index set $\{1,\dots, n\}$ 
  gives,
  \begin{equation*}
  T({\E L}_{\pi(1)}(x_{\pi(1)})\cdots {\E 
  L}_{\pi(n)}(x_{\pi(n)}))\equiv
  T({\E L}_{1}(x_{1})\cdots {\E L}_{n}(x_{n}))\ ,
  \end{equation*}
  for any $\pi\in\E{P}_{\{1,\dots,n\}}$, in the sense of distributions.
\end{properties}

This symmetry property corresponds to the fact 
that the time-ordered products are functional 
derivatives of the local $S$-matrix. 
More crucial is the following causality 
property, which follows from the 
Eqn.~(\ref{b});

\begin{properties}{3}
\item {\bf Causality}. Consider any set of points $(x_1,\dots, x_n)\in M^n$
  and any full partition of the set $\{1,\dots, n\}$ into two non empty 
  subsets $I$ and $I^c$ such that no point $x_i$ with $i\in I$ 
  is in the past of the points $x_j$ with $j\in I^c$, 
  i.e. $x_i\notin J^{-}(x_j)$
  for any $i\in I$ and $j\in I^c$. 
  Then the time-ordered distributions are required to satisfy the 
  following factorization property
  \begin{equation*}
  T({\E L}_{1}(x_{1}) \cdots {\E L}_{n}(x_{n}))=
  T(\prod_{i\in I}{\E L}_{i}(x_{i}))\ 
  T(\prod_{j\in I^c}{\E L}_{j}(x_{j}))\ .
  \end{equation*}
\end{properties}

In the Epstein and Glaser scheme on Minkowski space-time
one requires, in addition, translation covariance 
of the time-ordered products. If the free field 
is among the possible terms in the Lagrangian one  
can show that the time-ordered products are 
sums of pointwise products of Wick polynomials 
with translation invariant numerical 
distributions. (Such products exist due to Theorem 
0 of Epstein and Glaser, an easy proof of which follows 
from our microlocal characterization of the domain 
of definition of Wick products.)  Moreover, the 
condition,
\begin{equation}
        [T({\E L}_{1}(x_{1}) \cdots {\E 
        L}_{n}(x_{n})),\varphi(y)]=\sum_{i=1}^n
        T({\E L}_{1}(x_{1}) \cdots \frac{\partial {\E 
        L}_{i}}{\partial \varphi}(x_{i}) \cdots {\E 
        L}_{n}(x_{n}))E(x_{i},y)\ ,
        \label{E:subwick}
\end{equation}
fixes the coefficients to be vacuum expectation values of 
time-ordered products of those Wick polynomials which are 
of lower order w.r.t. the chosen
interacting Lagrangian (from now on, 
we shall call them {\it sub}-Wick polynomials), 
hence the problem is reduced to a problem for 
numerical distributions.
Unfortunately, in 
the case of a curved space-time, we 
have not yet determined the
class of fields which are relatively local to the scalar free field,
i.e., what is known in the literature 
as the Borchers' class \cite{B:es}. One also needs a 
replacement of the condition of translation 
covariance.
Our idea is to impose a condition on the time-ordered distributions 
which in 
a sense employs both ideas of invariance and spectrality crucial in the
Minkowskian case. Since, as emphasized in the previous section,
spectrality for us means wave front sets properties we now look for
a condition which fixes the properties of the singularities of the 
time-ordered distributions. We use the graph theoretic definitions of 
Section 2.

\begin{properties}{4}
\item {\bf Spectrality}. 
   For the expectation value $t_n\in\E{D}^{\prime}(M^n)$, $n\ge 2$, 
   of any time-ordered product it holds,
\begin{equation*}
\WF(t_{n})\subset\Gamma_{n}^{\timo}\ ,
\end{equation*}
\end{properties} 
where 
  \begin{equation*}
  \begin{split}
    \Gamma_{n}^{\timo}=
    \biggl\{&(x_{1},k_{1};\dots;x_{n},k_{n})\in 
    \udot{T}^{\ast} M^{n}\quad|\quad\exists\quad\text{a graph
    $G\in\C{G}_n$ and} \\
    &\text{an immersion $(x,\gamma,k)$ of $G$ in which  
         $k_e$ is future directed}\\ 
    &\text{whenever $x_{s(e)}\notin J^{-}(x_{r(e)})$ and such that,}\\
    &k_{i}=\sum_{m:s(m)=i} k_{m}(x_{i})-\sum_{n:r(n)=i}k_{n}(x_{i})
    \biggr\}\ .
  \end{split}
  \end{equation*}

This may be motivated by the fact that, for non coinciding points, 
$t_n$ can be expressed in terms of the usual Feynman graphs and for the set 
of coinciding points we have an infinitesimal remnant of translation 
invariance, since all covectors at coinciding points sum up to zero.

We can now formulate  a microlocal version of Theorem $0$ of Epstein and
Glaser. 
\begin{Thm} [Microlocal Theorem 0]   
If\ {\bf P4} holds for $t_n$ then
\begin{equation*}
  t_{n}(x_{1},\dots,x_{n})\multiwick{l_{1}}{x_{1}}{l_{n}}{x_{n}}\ ,
\end{equation*}
is a well defined operator-valued distribution, for any $n$ and any choice of
indices $l_{1}, \dots, l_{n}$, on the dense 
invariant domain $\C{D}$ in the Hilbert space $\C{H}_{\omega}$.
\end{Thm}
\begin{pf}
Let $\Psi\in\C{D}$. The vector-valued 
distribution 
$\multiwick{l_{1}}{x_{1}}{l_{n}}{x_{n}}\Psi$ is a restriction of
$\delta^{l} \Psi/\delta f^{l}$, $l=\sum_{i=1}^{n}l_{i}$, to a
partial diagonal with wave front set contained in 
$\bigcup_{x\in 
M^{n}}\{x\}\times{\overline{V}}_{-}^{\times n}$. Since 
$\Gamma_{n}^{\timo}$ does not contain elements of the 
form $(x_1,k_1,\dots,x_n,k_n)$ with $k_i\in{\overline{V}}_{+}$, $i=1,\dots,n$, 
the product is a well defined vector-valued 
distribution, and after smearing with some test 
function one obtains again a vector in  $\C{D}$. 
\end{pf} 

In particular, formula (\ref{E:wicktime}) 
makes sense everywhere, provided the expectation 
values of all time-ordered products of sub-Wick 
polynomials satisfy {\bf P4}. Moreover, every 
expansion into a sum of products of Wick 
polynomials by numerical distributions which satisfies 
(\ref{E:subwick}) is of this form: 
\begin{properties}{5}
\item {\bf Causal Wick Expansion}.
\begin{equation}\label{E:truewicktime}
\begin{split} 
T({\E L}_{1}(x_1)\cdots{\E L}_{}(x_n))=\sum_{j_1,\dots,j_n}
(\Omega_\omega,T({\E L}_{1}^{(j_{1})}&(x_1)\cdots {\E L}_{n}^{(j_{n})}(x_n))
\Omega_\omega)\times\\ 
&\times 
\frac{\multiwick{j_{1}}{x_1}{j_{n}}{x_n}}{j_{1}!\cdots
j_{n}!}\ .
\end{split}
\end{equation}
\end{properties}

\section{Inductive Construction up to the Small Diagonal}

The properties defined in the previous section allow us to set up an
inductive procedure in the spirit of Epstein and Glaser. 
We rely on a variation of their construction proposed by Stora
\cite{B:stora}. We start with a linear space ${\E W}$ of 
Wick polynomials which contains all respective sub-Wick 
polynomials and want to define the 
time-ordered products $T({\E L}_{1}(x_1)\cdots{\E L}_{}(x_n))$  
as a family of operator-valued distributions 
which are multilinear in the entries ${\E 
L}_{i}\in {\E W}$ and satisfy the properties {\bf P1-5}. 
We start the induction by setting $T({\boldsymbol 1})={\boldsymbol 1}$ 
and $T({\E L})={\E L}$  and assume that 
the time-ordered products for $1<l\le n-1$ factors have been
constructed and satisfy all the defining 
properties.  In a first step, we aim at constructing
time-ordered products of $n$ factors on $\meno{M^{n}}{\Delta_n}$ 
where $\Delta_n$ is the small diagonal
submanifold of $M^n$, i.e. the set of points 
$(x_1,\dots, x_n)$ with the property $x_1=x_2=\cdots=x_n$.  

We use the space-time notion of causality in order to define a certain 
partition of unity for $\meno{M^n}{\Delta_n}$:
Let us denote by $\C{J}$ the family of all non-empty proper subsets $I$ of 
the index set $\{1,\dots,n\}$ and define, accordingly, the sets
$\C{C}_{I}=\{(x_{1},\dots,x_{n})\in M^{n}\ |\ 
x_{i}\notin {J}^{-}(x_{j}), i\in I, j\in I^c \}$ for any $I\in\C{J}$. 
Note that the defining relation for the 
$\C{C}_{I}$'s  is related to causality on $M$ and {\it not} on $M^n$.
It is fairly easy to show that
\begin{Lemma} 
Let $M$ be a globally hyperbolic space-time, then it holds
 \begin{equation*}
  \bigcup_{I\in\C{J}}\C{C}_{I}= \meno{M^{n}}{\Delta_n}\ .
 \end{equation*}
\end{Lemma}
\begin{pf}
The inclusion $\cup_{I}\C{C}_{I}\subset\meno{M^{n}}{\Delta_n}$ is obvious. 
The opposite inclusion is proved as follows.
Consider any set of points $(x_1,\dots,x_n)$ such that 
$x_{i}\neq x_{j}$ for some $i\neq j$, then the points $x_i$ and $x_j$ 
can be separated by a Cauchy surface $\Sigma$ as follows from the 
global hyperbolicity assumption. One may choose it as to contain 
none of the points $x_k$, $k=1,\dots, n$. Hence, defining 
$I=\{k\ |\  x_k\in J^{+}(\Sigma)\}$ and noting that $I\in\C{J}$ we find
$(x_1,\dots,x_n)\in\C{C}_I$. 
\end{pf}

We use the short hand notations
\begin{equation}
        T^{I}(x_{I})=T(\prod_{i\in I}{\E 
        L}_{i}(x_{i})),\qquad x_{I}=(x_{i},i\in I)\ .
        \label{E:short}
\end{equation}
The first step now is to set on any $\C{C}_{I}$
\begin{equation}\label{E:tensor}
T_{I}(x)\doteq
T^{I}(x_{I})
\ T^{I^c}(x_{I^c})\ ,
\end{equation}
as an operator-valued distribution since
according to the induction hypothesis and the 
fact that $I$ is proper
this is a well defined operation on $\E{D}(\C{C}_{I})$. 

We now glue together all operators
previously defined on different elements of the cover. 
For this we need to prove a sheaf consistency condition. Indeed, 
different $\C{C}_I$'s overlap 
but due to the causality hypothesis {\bf P3} and 
the causal Wick expansion {\bf P5} valid for the 
lower order terms, the following property holds:
\begin{Prop} 
For any choice of $I_{1}, I_{2}\in\C{J}$ such that
$\C{C}_{I_1}\cap\C{C}_{I_2}\neq\emptyset$ we have
\begin{equation*}
  T_{I_{1}}\!\!\restriction_{\C{C}_{I_1}\cap\C{C}_{I_2}}
  =T_{I_{2}}\!\!\restriction_{\C{C}_{I_1}\cap\C{C}_{I_2}}\ ,
\end{equation*}
in the sense of operator-valued distributions over $\meno{M^{n}}{\Delta_n}$.
\end{Prop}
\begin{pf}Let $I_1,I_{2} \in\C{J}$ and 
$x=(x_1,\dots,x_n)\in\C{C}_{I_1}\cap\C{C}_{I_2}$.
Using the causality property {\bf P3} which is by 
assumption valid for time-ordered products of 
less than $n$ factors we find,
\begin{equation}\label{E:split}
\begin{split}
T^{I_1}(x_{I_{1}})=
T^{I_1\cap I_2}(x_{I_1\cap I_2} )
\ T^{I_1\cap I^c_2}(x_{I_1\cap I^c_2})\ ,\\
T^{I^c_1}(x_{I^c_1})
=T^{I^c_1\cap I_2}
(x_{I^c_1\cap I_2})
\ T^{I^c_1\cap I^c_2}(x_{I^c_1\cap I^c_2})\ ,
\end{split}
\end{equation}
and similarly for $T^{I_2}$ and $T^{I_2^c}$.
Now the terms $T^{I_1\cap I^c_2}$
and $T^{I^c_1\cap I_2}$ commute. Namely, they are based on mutually 
space-like points, thus using the Wick expansion for 
these terms this follows from  local commutativity
for the Wick polynomials of the free scalar field 
$\varphi$. Hence from definition~(\ref{E:tensor}) 
we get on $\C{C}_{I_1}\cap\C{C}_{I_2}$
\begin{alignat*}{2}
T_{I_1} 
&=T^{I_1\cap I_2} T^{I^c_1\cap I_2}
T^{I_1\cap I^c_2}T^{I^c_1\cap I^c_2}\ ,&
&\qquad(\text{Eqn.} (\ref{E:tensor})+\text{\bf P3}),\\
&=T^{I_2}T^{I^c_2}\ ,& &\qquad (\text{Eqn.} (\ref{E:split})),\\
&=T_{I_2}\ ,& 
&\qquad (\text{Eqn.} (\ref{E:tensor})).
\end{alignat*}
\end{pf}

Let now $\{f_{I}\}_{I\in \C{J}}$ be a locally finite smooth partition of 
unity of $\meno{M^{n}}{\Delta_n}$ subordinate to $\{\C{C}_{I}\}_{I\in \C{J}}$. 
We formally define, following (\ref{E:short}) and (\ref{E:tensor}),
\begin{equation}\label{E:pudef}
{^{0}T}({\E L}_{1}(x_{1})\dots{\E L}_{n}(x_{n}))\doteq
\sum_{I\in \C{J}} f_{I}\ T_{I}\ .
\end{equation}
Hence, we get our first crucial result, namely,
\begin{Thm} 
The symbols ${^{0}T}$ are well defined 
operator-valued distributions
over $\meno{M^{n}}{\Delta_n}$ which satisfy the defining properties 
on $\C{D}\subset \C{H}_\omega$.
\end{Thm}
\begin{pf}
We first prove that the definition does not depend on the choice of the
partition of unity. Indeed,
let $\{f_{I}^{\prime}\}_{I\in\C{J}}$ be another such partition. Consider
$x\in\meno{M^{n}}{\Delta_n}$, and let $\C{K}=\{I\in\C{J}\ |\ 
x\in\C{C}_{I}\}$. Then there 
exists a neighbourhood $V$ of $x$ such that 
$V\subset\cap_{I\in\C{K}}\C{C}_{I}$, 
and $\supp (f_{I})$ and $\supp (f^{\prime}_{I})$ do not meet $V$ for all
$I\notin\C{K}$. In this case
$$
\sum_{I\in\C{J}}(f_{I}-f^{\prime}_{I})\ T_{I}\!\!\restriction_{V}=
\sum_{I\in\C{K}}(f_{I}-f^{\prime}_{I})\ T_{I}\!\!\restriction_{V}.
$$
However, on $V$ $T_{I}$ is independent of the choice of $I\in\C{K}$. Since
$\sum_{I\in\C{K}}f_{I}=\sum_{I\in\C{K}}f^{\prime}_{I}=1$ on $V$, 
we arrive at the conclusion. Furthermore, an inspection of the formula
readily gives that the operator ${^0}T$ is  defined 
on the domain $\C{D}$, the microlocal domain of definition  of 
the Wick monomials, because of induction. Hence property {\bf P1}.

As far as the symmetry property {\bf P2} is concerned we 
just observe that the permuted distribution
${^{0}T}^{\pi}(x_1,\ldots,x_n)=
{^{0}T}({\E L}_{\pi(1)}(x_{\pi(1)}) \ldots {\E L}_{\pi(n)}(x_{\pi(n)}))$, 
has the expansion,
\begin{equation*}
{^{0}T}^{\pi}=\sum_{I\in 
J}f_{I}^{\pi}T_{I}^{\pi}=
\sum_{I\in J}f^{\pi}_{\pi(I)}\ T_{\pi(I)}^{\pi}\ ,
\end{equation*}
where we used the fact that the set $J$ is invariant under permutations, but 
$T_{\pi(I)}^{\pi}=T_{I}$ and $\{f_{\pi(I)}^{\pi}\}_{I\in J}$ is a partition 
of unity subordinate to $\{\C{C}_{I}\}_{I\in J}$, so symmetry follows from the 
result of the previous paragraph about the independence of ${^{0}T}$ 
on the choice of the partition of unity. 

Causality {\bf P3} follows from an argument similar to the one used for the 
independence from the partition of unity. Indeed, taken any point 
$x\in\meno{M^n}{\Delta_n}$, as before 
$x\in V\subset\cap_{I\in\C{K}}{\E C}_{I}$.
From (\ref{E:pudef}),
\begin{equation*}
{^0}T(x)=\sum_{I\in\C{K}} f_I (x) T_I (x)\ .
\end{equation*}
Since $T_I\!\!\restriction_{V}$ does not depend anymore on 
$I\in\C{K}$ and $\sum_{I\in\C{K}}f_I =1$ over $V$ hence 
${^0}T(x)\equiv T_I(x)$ which from (\ref{E:tensor}) 
satisfies causality by definition.

Now, we want to show that  property {\bf P4} 
holds on $\meno{M^n}{\Delta_n}$. It is sufficient 
to check that this property is satisfied for each 
$T_I$ on $\C{C}_{I}$. We apply 
Wick Theorem to the components of the product in the definition
(\ref{E:tensor}). It can be easily checked that the distributions 
$t_{I}$ in the Wick expansion of $T_I$ are sums of terms of 
the following form
\begin{equation}\label{E:tensorp}
f_{I}(x) t^{I}(x_{I}) t^{I^c}
 (x_{I^c})\cdot
 \prod_{(i, j)\in I\times I^c}\omega_{2}(x_i,x_j)^{a_{i,j}}\ ,
\end{equation}
with $a_{i,j}\in\Bbb{N}_{0}$, and where $t^{I},t^{I^c}$ 
are expectation values of lower order time-ordered 
products.

The wave front set of (\ref{E:tensorp}) is contained in the convex 
combination of the wave front sets of its factors. Hence it is 
given in terms of immersions of graphs  with vertex sets 
$I,I^c$, resp., and of $a_{i,j}$ graphs with vertex sets 
$\{i,j\}$. All these immersions satisfy the condition in 
{\bf P4}, the first two by assumption, and the last ones 
because of the definition of $\C{C}_{I}$ and the properties 
of the wave front set of a Hadamard state. 
The union of these graphs is a graph with 
vertices $\{1,\dots,n\}$, and any convex combination of the 
components is given by an admissible immersion of this graph.

Finally, property {\bf P5} follows from  expression~(\ref{E:tensor})
by a straightforward application of the (generalized) Wick Theorem. 
\end{pf}

\section{Steinmann Scaling Degree and the Extension of Distributions}

We now want to extend ${^{0}T}({\E L}_{1}(x_1) \ldots {\E L}_{n}(x_n))$ 
to the whole $M^n$. 
As discussed before the problem can be reduced to 
the extension of the numerical time-ordered distributions 
${^0}t(x_1,\ldots, x_n) \doteq (\Omega_{\omega},
{^{0}T}({\E L}_{1}(x_1) \ldots {\E L}_{n}(x_n))\Omega_{\omega})$. 

The extension can be performed 
in two steps. 
First ${^0}t$ is extended by continuity to the subspace of 
test-functions which vanish on $\Delta_n$ up to a certain 
order, and then it is arbitrarily defined on a 
complementary subspace. It is this last step 
which corresponds to the method of counterterms in the classical procedure of
perturbative renormalization. The extension of 
${^0}t$ by continuity requires some topology on test-function space. 
The seminorms used by Epstein and Glaser in their paper are quite 
complicated, and their generalization to curved space-times appears to be 
rather involved. 
We found it preferable therefore to apply a different method
already introduced by Steinmann \cite{B:steinmann}, namely the concept 
of {\it scaling degree} at a point of a distribution (see 
also \cite{B:dm}). Its generalization to 
curved space-time is very similar to the concept of the scaling limit as 
introduced by Haag, Narnhofer and Stein \cite{B:hns} 
and further developed by Fredenhagen and Haag 
\cite{B:fh}. A similar technique is used in 
\cite{B:scharf}.

On Minkowski space, by translation invariance, the 
distribution is in terms of relative coordinates 
everywhere defined up to the origin, and there the concept 
of the {\it scaling degree} at a point leads to a 
rather smooth and economic method of 
renormalization, see e.g. \cite{B:prange} where 
the relation to differential renormalization is 
elaborated. On a curved space-time one needs the 
corresponding notion for a scaling degree with 
respect to the submanifold $\Delta_n$, and one 
also needs some uniformity of the singularity 
along $\Delta_n$ as well as control of the wave 
front sets during the extension process.

Our strategy will be that, at first we introduce this improvement 
in the case of $\Bbb{R}^n$, then
we discuss the case of manifolds. There we try to set up a procedure which
allows to restrict the discussion to the pointwise case.

\subsection{The scaling degree}
\label{SS:scaling}

For simplicity, we work at first on $\Bbb{R}^d$. 
 Hence, consider a distribution $t\in\E{D}^{\prime}(\Bbb{R}^d)$.
Let the action of
the positive reals (dilations) be defined via the map 
\begin{alignat*}{2}
\Lambda : \Bbb{R}_{+}&\times
\E{D}(\Bbb{R}^d)\ &\longrightarrow\ &\ \E{D}(\Bbb{R}^d)\\ 
&(\lambda, \phi)            \ &\longmapsto\ &\ \phi^{\lambda}\doteq
\lambda^{-d}\phi(\lambda^{-1}\ \cdot\ )\ , 
\end{alignat*}
and obtain, by pull-back, the map over distributions 
$t\in\E{D}^{\prime}(\Bbb{R}^d)$ as,
\begin{equation*}
(\Lambda^{\ast}t)(\phi)\doteq t_{\lambda}(\phi)\doteq t(\phi^\lambda)\ ,
\end{equation*}
where this operation in 
case $t\in L^{1}_{\text{loc}}(\Bbb{R}^d)$ is given by the explicit formula,
\begin{equation}\label{E:formula}
t_\lambda (\phi) = \int t(\lambda x) \phi(x) d^{d}x\ , \qquad \forall \phi\in 
\E{D}(\Bbb{R}^d)\ .
\end{equation}
The map $\Lambda$ is clearly 
continuous w.r.t. the topology of $\E{D}(\Bbb{R}^d)$ and 
we shall sometimes use the previous formula~(\ref{E:formula}), 
by the usual abuse of notation, also in the general case. 

We say that $t$
has {\it scaling degree} $\sd(t)=\omega$ w.r.t. the origin in $\Bbb{R}^d$, if
$\omega$ is the infimum of all 
$\omega^\prime\in\Bbb{R}$ for which,
\begin{equation}\label{E:scaling}
\lim_{\lambda\downarrow 0}
\lambda^{\omega^{\prime}} t_{\lambda}=0\ ,
\end{equation}
holds in the sense of $\E{D}^{\prime}(\Bbb{R}^d)$.

It should be clear from the definition that every distribution  
$t\in\E{D}^{\prime}(\Bbb{R}^d)$ has a scaling 
degree $\omega\in[-\infty,+\infty[$. If the distribution is not defined
at the point we want to check, then the scaling degree might also be
equal to $+\infty$. We give some examples.

\begin{Exam}$\phantom{c}$\par
\begin{enumerate}{\rm
\item Trivial example. Every $\phi\in\E{E}(\Bbb{R}^d)$ has $\sd(\phi)\le 0$.
\item Dirac measure. Let 
$\mu\in\E{E}^{\prime}(\Bbb{R}^d)$ with
$\mu(\phi)=\phi(0)$, $\phi\in\E{E}(\Bbb{R}^d)$, 
then $\sd(\mu)=d$.
\item Feynman propagator. In the case of a free massive 
scalar field which is covariant under translation on Minkowski space-time, the 
Feynman propagator can be written as
\begin{equation*}
E_{F}(x) = (2\pi)^{-d} \int \frac{e^{ip\cdot x}}{p^2 - m^2 +i\epsilon} 
d^{d}p\ ,
\end{equation*}
from which it is readily seen that $\sd(E_{F})= d-2$.
\item Homogeneous distributions. If $t\in\E{D}^{\prime}(\Bbb{R}^d)$ is
homogeneous of order $\alpha$ at the origin, 
i.e. $t_\lambda = \lambda^{\alpha} t$, then $\sd(t)=-\alpha$.
\item Infinite degree. The smooth function $x\rightarrow \exp(1/x)$, $x\in
\Bbb{R}_+$, is not defined at the origin and its scaling degree w.r.t. the 
origin is clearly infinite.
}
\end{enumerate}
\end{Exam}

As inferred from the 4th example, the scaling degree may be seen as a 
generalization of the notion of the degree of homogeneity. 
Actually, our extension method is similar to the extension to all space of a 
homogeneous distribution as discussed 
in H\"ormander's book \cite{B:micro} which on the other hand is also
quite similar to the Epstein and Glaser procedure of distribution splitting 
\cite{B:eg}. The fact 
that  homogenous extensions not always exist is 
the mathematical origin of the logarithmic 
corrections to scaling found in renormalization.  
Here, a discussion about space-time symmetries and their 
implementation after renormalization 
is absent. It will be presented in \cite{B:bfrev}. 

\begin{Lemma}\label{L:propsd} The scaling degree obeys the following 
properties:
\begin{itemize}
\item[$(a)$] 
Let $t\in\E{D}^{\prime}(\Bbb{R}^d)$ have $\sd(t)=\omega$ at 
$0$ , then 
\begin{enumerate}
\item Let $\alpha\in\Bbb{N}^d$ be any multiindex, 
then $\sd(\partial^\alpha t)\le\omega +|\alpha|$\ .
\item Let  $\alpha\in\Bbb{N}^d$ be any multiindex,
then $\sd( x^\alpha t)\le\omega -|\alpha|$\ . 
\item Let $f \in \E{E}(\Bbb{R}^d)$, then 
$\sd(f t)\le\sd(t)$\ .
\end{enumerate}
\item[$(b)$] 
For 
$t_{i}\in\E{D}^{\prime}(\Bbb{R}^{d_{i}}),i=1,2$ 
we have $\sd(t_{1}\otimes 
t_{2})=\sd(t_{1})+\sd(t_{2})$\ .
\end{itemize}
\end{Lemma}
\begin{pf}
The first two cases in $(a)$ as well as $(b)$ are straightforward. 
The third case in $(a)$ follows from the fact that, by the 
Banach-Steinhaus principle, a convergent sequence 
of distributions is uniformly bounded. Hence, for 
every $\omega^{\prime}>\sd(t)$ and every compact 
set $K\subset\Bbb{R}^d$ there is some polynomial 
$P$ such that
 \begin{equation}
 |\lambda^{\omega^{\prime}}t_{\lambda}(\phi)|
 \le\sup_{x\in 
 K}|P(\partial)\phi(x)|\equiv||\phi||_{\infty,P}\ .     
 \label{E:Banach}       
 \end{equation}
 Hence, for $f\in\E{E}(\Bbb{R}^d)$, we have
 \begin{equation}
 |(ft)_{\lambda}(\phi)|=
 |t_{\lambda}(f_{\lambda}\phi)|
 \le\lambda^{\omega^{\prime}}||f_{\lambda}\phi||_{\infty,P}\ .  
 \end{equation}
 The statement follows now from the boundedness of the 
 sequence $||f_{\lambda}\phi||_{\infty,P}$ as $\lambda\rightarrow 0$.
\end{pf}

\subsection{Extensions of distributions to a point}

We now want to show how to extend a distribution 
$t\in\E{D}^{\prime}(\meno{\Bbb{R}^d}{\{0\}})$ to all space 
by using the concept of the scaling degree. 
The scaling degree can easily be defined for such 
distributions by restricting the test functions 
appropriately. Equivalently, we can also, for each $\chi 
\in\E{E}(\Bbb{R}^d)$ with $0\not\in\supp(\chi)$ look at the 
behaviour of the sequences $\chi t_{\lambda}$, now 
considered as sequences in $\E{D}^{\prime}(\Bbb{R}^d)$. 

There are three possible cases; when the scaling degree 
is $+\infty$; in this case no extension to a 
distribution on $\Bbb{R^d}$ exists; when the scaling 
degree $\omega$ is finite, but $\omega\ge d$; then 
a finite dimensional set of extensions exists; or 
otherwise $\omega< d$. We first study the third case. 

\begin{Thm} \label{T:noren}
Let $t_0\in\E{D}^{\prime}(\meno{\Bbb{R}^d}{\{0\}})$ have scaling 
degree $\omega< d$ w.r.t. the origin. There exists a unique 
$t\in\E{D}^{\prime}(\Bbb{R}^d)$  with scaling degree $\omega$ such that
$t(\phi)=t_0(\phi)$, $\phi\in \E{D}(\meno{\Bbb{R}^d}{\{0\}})$.
\end{Thm}
\begin{pf}
The uniqueness is easy. Indeed, the difference among two possible extensions
would be a distribution with support at $\{0\}$. By a well known structural
Theorem of distribution theory this last is given by $P(\partial)\delta$
where $P$ is a polynomial of degree $\text{deg}(P)$ and $\delta$ is Dirac
measure at the origin. But this distribution has 
scaling degree equal to $d+\text{deg}(P)$, hence a contradiction.

Let us now consider a smooth function of compact support $\vartheta$ such
that $\vartheta=1$ in a neighbourhood of the origin. Set 
$\vartheta_{\lambda}(x)\doteq\vartheta(\lambda x), 
\lambda\in\Bbb{R}$ and
\begin{equation*}
t^{(n)}\doteq (1-\vartheta_{2^n})t_{0}\ , \qquad n\in\Bbb{N}\ ,
\end{equation*}
where now $t^{(n)}$ is a sequence of distributions defined on the 
whole $\Bbb{R}^d$. We wish to show that the sequence converges in the
weak$^\ast$ topology of $\E{D}^{\prime}(\Bbb{R}^d)$. 
Because of the sequential completeness of 
$\E{D}^{\prime}(\Bbb{R}^d)$ it is sufficient to prove 
that it is a Cauchy sequence. Let 
$\phi\in\E{D}(\Bbb{R}^d)$ and look at
\begin{equation}\label{E:cauchyseq}
\begin{split}
(t^{(n+1)} - t^{(n)})(\phi) &= 
(\phi t_{0})(\vartheta_{2^{n}}-\vartheta_{2^{n+1}}) \\
&=2^{-nd}(\phi t_{0})_{2^{-n}}(\vartheta-\vartheta_{2})\ .
\end{split}
\end{equation}
According to Lemma~\ref{L:propsd}, $(a) 3.$, this sequence is 
majorized, for every $\omega^\prime\in\ ]\omega,d[$ by 
$\text{const.}\ 2^{n(\omega^\prime -d)}$, hence it is 
summable as required.
The limit
\begin{equation*}
t(\phi)\doteq \lim_{n\to\infty} t^{(n)}(\phi)\ ,\qquad\forall 
\phi\in\E{D}(\Bbb{R}^d)\ ,
\end{equation*}
then defines an extension of $t_{0}$. It is 
obvious that the scaling degree of $t$ is not 
smaller than $\omega$.
It remains to proof that it
is not bigger than $\omega$.

Pick $\phi\in\E{D}(\Bbb{R}^d)$ and consider the following expression:
\begin{equation*}
t_{\lambda}(\phi) =
\lim_{n\to\infty}
\lambda^{ -d}t_{0}((1-\vartheta_{2^n})\phi_{\lambda^{-1}})\ .
\end{equation*}
Let $R,\epsilon >0$ be such that $\supp(\phi)\subset\{x, |x|< R\}$
and $\vartheta(x)=1$ for $|x| <\epsilon$. Then,
\begin{equation*}
(1-\vartheta(2^n x))\phi(\lambda^{-1}x)=0\ ,
\end{equation*}
whenever $2^{-n}\epsilon > \lambda R$. Let us choose 
$n_{\lambda}\in\Bbb{N}$
such that 
$2^{-n_{\lambda}}\epsilon > \lambda R > 2^{-(n_{\lambda}+1)}\epsilon$.
 
We have,
\begin{equation}\label{E:limit}
\begin{split}
t_{\lambda}(\phi)&=
\sum_{n=n_{\lambda}}^{\infty}\lambda^{-d} 
t_{0}((\vartheta_{2^{n}}-\vartheta_{2^{n+1}})\phi_{\lambda^{-1}})\\
&=\sum_{n=n_{\lambda}}^{\infty}(2^{n}\lambda)^{-d}
(t_{0})_{2^{-n}}((\vartheta-\vartheta_{2})\phi_{2^{-n}\lambda^{-1}})\ .
\end{split}
\end{equation}

The set 
$\{(\vartheta-\vartheta_{2})\phi_{\mu},\mu<\text{const.}\}$ 
is bounded in $\E{D}(\meno{\Bbb{R}^d}{\{0\}})$.
Hence for every
$\omega^{\prime} > \omega$
we find a constant $c>0$ such that
\begin{equation*}
\left| (t_{0})_{2^{-n}}((\vartheta-\vartheta_{2})
\phi_{2^{-n}\lambda^{-1}})\right|\le\ c\  
2^{n\omega^{\prime}},\qquad n\ge n_{\lambda}\ .
\end{equation*}
Inserting this estimate back to Eqn.~(\ref{E:limit}) we have 
\begin{equation*}
\begin{split}
\left|t_{\lambda}(\phi) \right| &<
\ c\ \lambda^{ -d} \sum_{n=n_{\lambda}}^{\infty}
2^{-n(d-\omega^{\prime})}= \ c\  
\lambda^{-d}\frac{2^{-n_{\lambda}(d-\omega^{\prime})}}
{1-2^{-(d-\omega^{\prime})}}\\
&\le\frac{c}{1-2^{-(d-\omega^{\prime})}}
\lambda^{-d} \left (\frac{2R}{\epsilon}
\right )^{d-\omega^{\prime}}
\lambda^{d-\omega^{\prime}}\le c^\prime 
\lambda^{ -\omega^{\prime}}
\end{split}
\end{equation*}
for some constant $c^\prime >0$. 
This proves the assertion.
\end{pf}

We now deal with the extension
procedure in case a distribution has a finite scaling degree $\omega\ge d$. 
This
extension procedure corresponds to 
renormalization in other schemes.
To adhere more to the standard notation we 
introduce the degree of singularity $\rho\doteq\omega-d$.  
This is the analog of the  degree of divergence of a 
Feynman diagram. 

Let $\E{D}_{\rho}(\Bbb{R}^d)$ be the set
of all smooth functions of compact support which vanish of order 
$\rho$ at the origin and let $W$ be a projection 
from $\E{D}(\Bbb{R}^d)$ onto 
$\E{D}_{\rho}(\Bbb{R}^d)$. Since the orthogonal 
complement of $\E{D}_{\rho}(\Bbb{R}^d)$ 
consists of the derivatives of the 
$\delta$-function up to order $\rho$, $W$ is of 
the form
\begin{equation}
        W\phi=\phi-\sum_{|\alpha|\le\rho}\fra{w}_{\alpha}
        \partial^{\alpha}\phi(0)\ ,
        \label{E:W-operation}
\end{equation}
with
$\fra{w}_{\alpha}$ being  smooth functions of compact support such that 
$\partial^{\alpha}\fra{w}_{\beta}(0)=\delta^{\alpha}_{\beta}$. 

\begin{Thm}\label{T:reno} 
Let $t_0\in\E{D}^{\prime}(\meno{\Bbb{R}^d}{\{0\}})$ 
have a finite scaling
degree $\omega\ge d$. Then there exist extensions  
$t\in\E{D}^{\prime}(\Bbb{R}^d)$ of $t_{0}$ with the same scaling 
degree, and, given $W$, they are uniquely determined by their 
values on the test functions $\fra{w}_{\alpha}$. 
\end{Thm}
\begin{pf} 
Any $\phi\in\E{D}(\Bbb{R}^d)$ can be uniquely decomposed as
$\phi=\phi_1 +\phi_2$ where $\phi_1=\sum_{|\alpha|\le\rho}\fra{w}_{\alpha}
 \partial^{\alpha}\phi(0)$ and
$\phi_2\in\E{D}_{\rho}(\Bbb{R}^d)$. 
$\phi_2$ has the form,
\begin{equation*}
\phi_2(x)=\sum_{|\alpha|=[\rho]+1} x^{\alpha}\psi_{\alpha}(x)\ ,
\end{equation*}
with $\psi_{\alpha}\in\E{D}(\Bbb{R}^d)$. 
We set
\begin{equation}\label{E:renorm}
\langle t,\phi\rangle =\sum_{|\alpha|=[\rho]+1}\langle x^{\alpha} t_{0}, 
\psi_{\alpha}\rangle + \langle t,\phi_1 \rangle\ .
\end{equation}
Since, by Lemma~\ref{L:propsd}, $x^\alpha t_{0}$ has scaling degree equal
to $\rho-[\rho]-1+d$ which is strictly smaller than $d$ this term
has a unique extension by Theorem~\ref{T:noren}.

We now prove that $t$ has the same scaling degree as $t_0$. We write,
\begin{equation*}
t_{\lambda}(\phi) =
(t_{0}\circ  W)_{\lambda}(\phi)-
 \sum_{|\alpha|\le\rho}t(\fra{w}_{\alpha})
 \left(\partial^{\alpha}\phi\right)(0)
\lambda^{-d-|\alpha|}\ .
\end{equation*} 
The second term clearly has scaling degree less or equal 
to $\rho+d=\omega$. The first term can be written in the form
\begin{equation*}
(t_{0})_{\lambda}(W\phi)+
(t_{0})_{\lambda}((W\phi_{\lambda^{-1}})_{\lambda}-W\phi)\ .
\end{equation*}
By assumption, the first term has scaling degree $\omega$. 
To analyze the second term we write
\begin{equation*}
((W\phi_{\lambda^{-1}})_{\lambda}-W\phi)(x)=\sum_{|\alpha|\le\rho}
\left(\fra{w}_{\alpha}(x)-\lambda^{-|\alpha|}\fra{w}_{\alpha}(\lambda x)
\right)
\left(\partial^{\alpha}\phi\right)(0)\ .
\end{equation*}
(Note that $(\fra{w}_{\alpha}(\cdot)-
\lambda^{-|\alpha|}\fra{w}_{\alpha}(\lambda\ 
\cdot))\in\E{D}_{\rho}(\Bbb{R}^d)$). Using the  identity,
\begin{equation}\label{E:trivide}
\begin{split}
(\fra{w}_{\alpha}(x)-\lambda^{-|\alpha|}\fra{w}_{\alpha}(\lambda x))&=
\int_{\lambda}^{1} \frac{d}{d\mu}\mu^{-|\alpha|}\fra{w}_{\alpha}(\mu x)
\ d\mu\\
&=\int_{\lambda}^{1}\left( \mu x_k (\partial_k 
\fra{w}_{\alpha})(\mu x)-|\alpha|\fra{w}_{\alpha}(\mu 
x)\right)\mu^{-|\alpha|-1} \ d\mu\ ,
\end{split}
\end{equation}
we get, after a moment of reflection for the
exchange of the order between integration and duality, that,
\begin{equation*}
\begin{split}
(t_{0})_{\lambda}(&(W\phi_{\lambda^{-1}})_{\lambda}-W\phi)=\\
&\sum_{|\alpha|\le\rho}
\left(\partial^{\alpha}\phi\right)(0)
\int_{\lambda}^{1} \mu^{-d-|\alpha|-1}
(t_{0})_{\lambda\mu^{-1}} 
((x_k \partial_k-|\alpha|)\fra{w}_{\alpha}) d\mu\ .
\end{split}
\end{equation*}
The  integrand can be estimated according to Lemma~\ref{L:propsd}. 
Indeed, for any $\omega^{\prime}>\omega$ we have,
\begin{equation*}
\left|(t_{0})_{\lambda\mu^{-1}}\left((x_k 
\partial_k-|\alpha|)\fra{w}_{\alpha}\right)\
\right| \le \text{const.} \left(\lambda^{-1}\mu 
\right)^{\omega^{\prime}}\ ,
\end{equation*}
and therefore
\begin{equation*}
\left|\int_{\lambda}^{1} \mu^{-d-|\alpha|-1}
(t_{0})_{\lambda\mu^{-1}}\left(  
(x_k \partial_k-|\alpha|)\fra{w}_{\alpha}\right) d\mu \right|
\le\text{const.}\ 
\lambda^{-\omega^{\prime}}\
\frac{1- \lambda^{\omega^{\prime}-d-|\alpha|}}{\omega^{\prime}-d -|\alpha|} 
\ ,
\end{equation*}
which proves the assertion.
\end{pf}

The expert reader can now proceed from this point to study the 
renormalizability of any theory 
which admits space-time translation covariance.
The ambiguity of the extension is given by terms localized over the origin.
The coefficients of these terms can be fixed by additional requirements, as
customary in perturbative quantum field theory. We refer the reader to
\cite{B:bfrev} for more details.

During this process, one needs estimates on the scaling degrees 
of the arising distributions, corresponding to the power 
counting rules. In addition to Lemma 
\ref{L:propsd}  estimates on scaling degrees of 
products of distributions (provided they exist) are required. These can be 
obtained by explicit calculations (see e.g. the analogous estimates in 
\cite{B:eg}). Much more elegant is a general method which 
exploits a microlocal version of the scaling degree. This 
technique is actually necessary if one wants to generalize 
the methods above to generic manifolds. We shall describe it 
in the next section.

\section{Surfaces of Uniform Singularity and the 
Microlocal Scaling Degree}

The generalization of the previous procedure to the case of submanifolds
is what we really need in the treatment of perturbation theory on curved 
spaces. Indeed, the description given in Section 4 led to
the notion of a scaling degree w.r.t. the small diagonal
$\Delta_n$ of the topological product $M^n$. Here
we classify the behaviour of distributions near some surface by a microlocal
version of the scaling degree. We introduce two different notions. The first
one, the (microlocal) scaling degree at a surface, involves only the 
surface under consideration, the second one, the transversal scaling degree, 
involves a fibration of the surface by transversal surfaces. The first notion
behaves very nicely under tensor products and restrictions, whereas the second
one admits an easy generalization of the extension procedure. As a 
matter of fact, the notions can be shown to be equivalent.

\subsection{Scaling degrees at submanifolds}

Let $\E{M}$ be a smooth paracompact manifold of dimension $d$ and
$t$ be a distribution in
$\E{D}^{\prime}(\E{M})$. Let $\E{N}\subset\E{M}$ be a submanifold such 
that the wave front set of $t$ is orthogonal to the tangent bundle $T\E{N}$
of $\E{N}$, i.e. for $(x,k)\in\WF(t)$ with 
$k\in T^{\ast}_{x}\E{M}, x\in\E{N}$,
\begin{equation}
\langle k,\xi\rangle = 0\ ,\qquad\forall\xi\in T_{x}\E{N}\ . 
\end{equation}
Under these circumstances, $t$ can be restricted to a sufficiently small
submanifold $\E{C}\subset\E{M}$ which intersects $\E{N}$ in a single point
$x_0$,  such that the intersection of their tangent spaces at $x_0$
is trivial and their sum spans the whole tangent space 
(the submanifolds $\E{C}$
and $\E{N}$ are {\it transversal}, see e.g. \cite{B:gp},
symbolically $\E{C}\pitchfork\E{N}$).
This is due to the fact that $\WF(t)$ does not intersect the conormal bundle
$N^{\ast}\E{C} =\{(x,k)\in T^{\ast}\E{M}| \langle k,\xi\rangle=0, 
\forall\xi\in T_{x}\E{C}\}$ of $\E{C}$. Namely, for 
$k\in T^{\ast}_{x_{0}}\E{M}$, 
$(x_0,k)\in\WF(t)$ we have $\langle k,\xi\rangle =0$ for 
$\xi\in T_{x_{0}}\E{N}$, hence $\langle k,\xi\rangle\neq 0$ for some
$\xi\in T_{x_{0}}\E{C}$, thus $(x_0,k)\notin N^{\ast}\E{C}$. But 
$\WF(t)\cap N^{\ast}\E{C}$ is a closed conical set in 
$\udot{T}^{\ast}_{\E{C}}\E{M}$, hence its complement is an open conical
neighbourhood of $\udot{T}^{\ast}_{x_{0}}\E{M}$, in particular it
contains a set $\udot{T}^{\ast}_{U_{0}}\E{M}$ where $U_{0}$ is an open
neighbourhood of $x_{0}$ in $\E{C}$. By choosing $\E{C}= U_{0}$ we arrive at 
the conclusion. So we proved,

\begin{Lemma}\label{L:restriction}
Let $t\in\E{D}^{\prime}(\E{M})$ be a distribution on a smooth manifold
$\E{M}$ and let  $\E{N}$ be a submanifold such that ${\rm\WF}(t)\perp
T\E{N}$. Then $t$ can be restricted to every sufficiently small 
submanifold $\E{C}$ such that $\E{N}\pitchfork\E{C}$ .
\end{Lemma}

The singularity of $t\!\!\restriction_{\E{C}}$ at $x_{0}$ may be classified
by a covariant extension of the notion of the scaling degree, or better by 
a slight extension which uses microlocal analysis. For the economy of the 
presentation we first look at the concept
of scaling degree at some surface $\E{N}$
which reduces for each transversal surface $\E{C}$ to the
scaling degree at the intersection point. This last will just be a pointwise
reduction of the general case we proceed to discuss right now.

Let $U$ be a star-shaped neighbourhood of the zero section 
$Z(T_{\E{N}}\E{M})$ and consider a map $\alpha : U\to \alpha(U)\subset
\E{N}\times\E{M}$ which is a diffeomorphism onto its range and such that
the following properties hold true
\begin{itemize}
\item[$(i)$] $\alpha(x,0)=(x,x)$, $x\in\E{N}$;
\item[$(ii)$]$\alpha(T\E{N}\cap U)\subset\E{N}\times\E{N}$;
\item[$(iii)$]$\alpha(x,\xi)\in\{x\}\times\E{M}$, $x\in\E{N}, 
\xi\in T_{x}\E{M}$;
\item[$(iv)$]$d_{\xi}\alpha(x,\cdot)\!\!\restriction_{\xi=0} = 
\id_{T_{x}\E{M}}$.
\end{itemize}

A concrete example of such a map $\alpha$ can be defined, whenever we
consider the manifold $\E{M}$ endowed with a (semi-)Riemannian metric, 
in terms of the exponential map, namely, 
$\alpha(x,\xi)\doteq (x,\exp_{x}\xi)$, provided the 
submanifold $\E{N}$ is totally geodesic, as will be the case in our 
applications. In the general case, we shall call the set of all such maps
by $\E{Z}$.

Let $\alpha\in\E{Z}$ and set $t^{\alpha} = ({\bf 1}\otimes t)\circ \alpha$ on 
$\E{D}^{\prime}(U)$ and $t_{\lambda}^{\alpha} (x,\xi)\doteq 
t^{\alpha}(x,\lambda\xi)$, $0<\lambda\le 1$. Here, 
$\langle {\bf 1}\otimes t, \phi\otimes\psi\rangle=\int\phi\cdot 
\langle t,\psi\rangle$ for test-densities $\phi\in\E{D}_{1}(\E{N})$ and
$\psi\in\E{D}_{1}(\E{M})$. Since $U$ is starshaped, $\lambda^{-1} U\supset U$
for $0<\lambda \le 1$, hence $t^{\alpha}_{\lambda}$ can be considered as a 
distribution on $\E{D}_{1}(U)$.

As a preliminary step we have the following 
\begin{Prop}\label{P:wfstalpha}
For any $t\in\E{D}^{\prime}(\E{M})$ which satisfies the hypothesis
of Lemma~\ref{L:restriction}, there exists a closed conic set
$\Gamma\subset \udot{T}^{\ast}U$ such that
\begin{itemize}
\item[$(i)$] $\Gamma\perp T(T\E{N}\cap U)$;
\item[$(ii)$]${\rm\WF}(t^{\alpha}_{\lambda})\subset\Gamma$.
\end{itemize}
\end{Prop}
\begin{pf} Since $\alpha$ maps $T\E{N}\cap U$ into $\E{N}\times\E{N}$, its
derivative $\alpha_{\ast} :TU\to T(\E{N}\times\E{M})$ maps $T(T\E{N}\cap U)$
into $T(\E{N}\times\E{N})$. But $\WF(t)\perp T\E{N}$ implies 
$\WF({\bf 1}\otimes t) \perp T(\E{N}\times\E{N})$, hence
\begin{equation*}
\WF(t^{\alpha})=\alpha^{\ast}
\WF(({\bf 1}\otimes t)\!\!\restriction_{\alpha(U)})\perp
\alpha^{-1}_{\ast}(T_{\alpha(U)}(\E{N}\times\E{N}))
= T(T\E{N}\cap U)\ .
\end{equation*}
Now, 
\begin{equation*}
\WF(t^{\alpha}_{\lambda})=\{(x,\xi;k)\in T^{\ast}_{(x,\xi)}(U)| 
(x,\lambda\xi; k)\in\WF(t^{\alpha})\}\ .
\end{equation*}
Here, we identified the cotangent spaces at the points $(x,\xi)$ and 
$(x,\lambda\xi)$ by the isomorphism induced by the 
diffeomorphism 
$U\to\lambda U$, $(x,\xi^{\prime})\to (x,\lambda\xi^{\prime})$, 
$\xi^\prime \in T_{x}\E{M}$. Now, let $\xi\in T_{x}\E{N}$ and 
$\eta \in T_{(x,\xi)}\E{N}$. We may identify $\eta$ with a vector in
$T_{(x,\lambda\xi)}\E{N}$ and observe that it is orthogonal to 
$\WF(t^{\alpha})$ and hence also to $\WF(t^{\alpha}_{\lambda})$. We now set
$\Gamma=\overline{\cup_{0<\lambda\le 1}\WF(t^{\alpha}_{\lambda})}$, where the
closure is performed within $\udot{T}^{\ast}U$. It remains to prove
$(i)$.

Let $(x,\xi_{n};k_{n})\in\WF(t_{\lambda_{n}}^{\alpha})$ be a convergent 
sequence in $T^{\ast}U$ with limit $(x,\xi;k)$, $k\neq 0$ and 
$\xi\in T_{x}\E{N}$. There is a corresponding sequence 
$(x,\lambda_{n}\xi_{n};k_{n})\in\WF(t^{\alpha})$. Let $\lambda\in [0,1]$ be 
a limit point of the bounded sequence $\{\lambda_{n}\}_{n\in\Bbb{N}}$. Then,
there is a subsequence converging to 
$(x,\lambda\xi;k)\in\WF(t^{\alpha})$. But $\lambda\xi \in T_{x}\E{N}$, hence
we have $k\perp T_{(x,\lambda\xi)}(T\E{N})$. If we again
identify the tangent spaces at $(x,\lambda\xi)$ and $(x,\xi)$ we obtain
the desired result.
\end{pf}

Choosing first any map $\alpha\in\E{Z}$ we are ready for the following:

\begin{Def}\label{D:msdalpha}
A distribution 
$t\in\E{D}^{\prime}(\E{M})$ has the microlocal 
scaling degree $\omega$ at a submanifold $\E{N}$
w.r.t. a closed conical set
$\Gamma_{0}\subseteq \udot{N}^{\ast}\E{N}$, symbolically 
$\omega=\msd_{\E{N}}^{\Gamma_0}(t,\alpha)$, if,
\begin{itemize}
\item[($i$)]  there exists a closed conic set 
$\Gamma\subset \udot{T}^{\ast}(T_{\E{N}}\E{M})$ with the properties
stated in  Proposition~\ref{P:wfstalpha}, with the first one replaced 
by $\Gamma\!\!\restriction_{T\E{N}}\subset 
\alpha^\ast(Z(T^\ast\E{N})\times\Gamma_0)$, 

\item[($ii$)] $\omega$ is the infimum of all 
those $\omega^\prime$ for which,
\begin{equation}\label{E:defmsd}
\lim_{\lambda\downarrow 0}\lambda^{\omega^{\prime}} 
  t^{\alpha}_{\lambda}=0\ ,
\end{equation}
in the sense of the H\"ormander 
topology on 
$\E{D}^{\prime}_{\Gamma}(T_{\E{N}}\E{M})$. 

\end{itemize}
\end{Def}

Now, depending on the position of $\Gamma_0$ one can give different refined
versions of the microlocal scaling degree. Indeed, when the inclusion in 
$\udot{N}^{\ast}\E{N}$ is proper we speak of the {\it strict} 
microlocal scaling degree. 
When $\Gamma_{0}\equiv \udot{N}^{\ast}\E{N}$, we 
call it simply the 
scaling degree at the submanifold, symbolically $\sd_{\E{N}}(t)$. 
Moreover, when 
$\Gamma_0 =\emptyset$ we speak of the {\it smooth} scaling degree.

The definition seems to depend on the choice of the map
$\alpha\in\E{Z}$. In our concrete 
case we could make use of the metric to choose a canonical diffeomorphism
$\alpha$ in terms of the exponential 
map, but for reasons which will become clear in the following it is helpful 
to proof its independence.

Before coming to that point we show an 
example for the computation of the scaling degree which is relevant 
for the physical discussion, namely, 
the generalization of the example $(3)$ in 
subSection \ref{SS:scaling}, the Feynman propagator $E_F$, 
to curved space-time. The Feynman propagator is considered as a distribution 
on $M\times M$, and we are interested in the 
microlocal scaling degree at the diagonal $\Delta_2\subset M\times M$. 
Indeed, the wave front set of $E_F$
is orthogonal to the tangent bundle 
of the diagonal. 
We choose $\alpha:T_{\Delta_2}M^2\rightarrow\Delta_2\times M^2\simeq
M\times M^2$ as $\alpha(x,\xi_1,\xi_2)=(x,\exp_x\xi_1,\exp_x\xi_2)$ and
obtain as on Minkowski space $\sd_{\Delta_2}(E_F)=d-2$.
A similar result holds for the 2-point Wightman distribution
$\omega_2$. So,
\begin{Lemma}\label{L:cmsdfp}
The  microlocal scaling degrees of the 
2-point Wightman distribution $\omega_2$ at the diagonal 
$\Delta_2$ w.r.t. $\Gamma_{0}=\{(x,k;x,-k)|x\in 
M,k\in\partial V_{+},k\neq 0\}$ 
is given by $\msd^{\Gamma_0}_{\Delta_2}(\omega_2)=d-2$,
where $d$ is the dimension of the space-time.
\end{Lemma}

\subsection{Invariance and properties for the scaling degrees}
Let us then choose two maps $\alpha_1, \alpha_2\in\E{Z}$
and state the following:
 
\begin{Prop} Let $t\in\E{D}^{\prime}(\E{M})$. 
Let $\omega_i=\msd_{\E{N}}^{\Gamma_0}(t,\alpha_i)$, $i=1,2$ be the 
microlocal scaling degrees 
w.r.t. $\E{N}$ and $\Gamma_0$ resp. for the two arbitrarily chosen maps. 
Then $\omega_1=\omega_2$.
\end{Prop}

\begin{pf} 
It is simple to check that
\begin{equation}\label{E:firste}
t^{\alpha_2}_{\lambda}(\phi)=
t_{\lambda}^{\alpha_1}(\phi\circ\beta_{\lambda}^{-1})\ ,\qquad\forall 
\phi\in\E{D}_1(T_{\E{N}}\E{M})\ ,
\end{equation}
where $\beta_{\lambda}(x,\xi)=\lambda^{-1}\beta(x,\lambda\xi)$ 
and $\beta=\alpha_{1}^{-1}\circ\alpha_{2}$.
Now, assume $\omega_1$ is the scaling degree for $t$ w.r.t. $\E{N}$ and 
$\Gamma_0$ associated with $\alpha_1$. We 
should proof that Eqn.~(\ref{E:defmsd}) for $\alpha_2$ converges in the 
sense of $\E{D}^\prime (T_{\E{N}}\E{M})$ as well, at the same rate as 
$\lambda\downarrow 0$. 

The convergence in the sense of distributions is simple. Indeed, 
if $\supp(\phi)\subset K$, $K$ a compact subset, 
then there exists a $\lambda_0$ such that
$\beta_\lambda (\supp(\phi))\subset K$ for all $\lambda\le\lambda_0$.
Hence it suffices by Banach-Steinhaus principle to proof that the family
$\{\phi\circ\beta_\lambda^{-1} | 0<\lambda\le\lambda_0\}$ is bounded, 
uniformly in 
$\lambda$, in $\E{D}_1 (K)$ w.r.t. the family of continuous seminorms 
which gives the appropriate Fr\'echet topology. This check proceeds 
easily from the chain rule and 
the verifiable fact that the only contribution 
comes from the $0$-th and $1$-st order derivatives w.r.t. 
$\beta_\lambda^{-1}$. In the limit they are the only terms
which survive giving resp. the identity map on the bundle and the 
derivative of the identity map.  Hence, we get the same rate of convergence
as far as plain distribution convergence is concerned.

A little bit trickier is the convergence in the sense of seminorms for
{\bf M2} ($b$). Starting again from Eqn.~(\ref{E:firste}),
via the multiplication of a smooth test function of compact support 
 $\psi$ such that $\psi
\equiv 1$ on a small neighbourhood of $\supp(\phi)$, 
we have that $\psi t$ is of compact support and then, by a 
partition of unity with functions with support on charts, that we are  
working on $\Bbb{R}^{\delta}\times\Bbb{R}^{d}$, where $\delta$ 
is the dimension of the submanifold $\E{N}$. 
Now, let us multiply the test function $\phi$  with the term 
$\exp(i\langle k,\ \cdot\ \rangle)$, and use inverse Fourier transform to get,
\begin{equation*}
\begin{split}
\widehat{\phi t^{\alpha_2}_\lambda}(k) &= 
{\beta_{\lambda}}^{\!\ast} t^{\alpha_1}_\lambda
(\phi\ {\exp(i\langle k,\ \cdot\ \rangle)})\\
&= \int  \widehat{\psi t_\lambda^{\alpha_1}}(p) 
I_\phi (p,k;\beta_\lambda ) d^{d+\delta}p\ ,
\end{split}
\end{equation*}
where,
\begin{equation*}
I_\phi (p,k;\beta_\lambda ) = \int e^{-i(\langle \beta_\lambda (\xi), p\rangle -
\langle\xi, k\rangle)}
\phi(\xi)\ d^{d+\delta} \xi\ ,
\end{equation*}
where in all these expressions the coordinates $\xi$ are the local 
coordinates of $T_{\E{N}}\E{M}$  
and $k$ and $p$ are their dual coordinates. 

We use the idea of the proof for the stationary phase
Theorem, see for instance Theorem 7.7.1 in H\"ormander books \cite{B:micro}.
Because of $\beta_{\lambda}\rightarrow \mbox{id}$ for 
$\lambda \rightarrow  0$ the oscillatory integral $I_{\phi}$ 
falls off rapidly outside of any conical neighbourhood  
of the diagonal $p=k$ in 
$\Bbb{R}^{d+\delta}\times\Bbb{R}^{d+\delta}$, uniformly 
for $\lambda$ sufficiently small, i.e. for every 
$\epsilon>0$ there exists a $\lambda_{0}>0$ such that for 
every $N\in\Bbb{N}$
\begin{equation}
        \sup_{0<\lambda<\lambda_{0}}\sup_{|p-k|>\epsilon|k|}
        (1+|p|+|k|)^{N}|I_{\phi}(p,k,\beta_{\lambda})|<\infty\ .
        \label{eq:oscint}
\end{equation}
Now let $\Gamma\subset\Bbb{R}^{d+\delta}$ be a closed cone 
such that
\begin{equation}
        \sup_{C}(1+|p|)^{N}|\widehat{\psi 
        t_\lambda^{\alpha_1}}(p)|\lambda^{\omega}\rightarrow 
        0\ ,\ \lambda \rightarrow 0\ ,
        \label{eq:decay1}
\end{equation}
for all closed cones $C$ with $C\cap\Gamma =\emptyset$ and 
all $N\in\Bbb{N}$. 

We now want to show that the same property 
holds for $\widehat{\phi t_{\lambda}^{\alpha_2}}$. So let 
$C$ be a closed cone such that the closed cone
\begin{equation}
        C'=\{p\in\Bbb{R}^{d+\delta},|p-k|\le \epsilon |k| \mbox{ for 
        some }k\in C\}
        \label{eq:decay2}
\end{equation}
does not intersect $\Gamma$. Then we split the region of 
integration over $p$ into the parts $|p-k|\le \epsilon |k|$ 
and the rest. In the first region we can estimate $k$ by 
$p$ and use the fast decay of 
$(1+|p|)^{N}|\widehat{\psi t_\lambda^{\alpha_1}}(p)|$ 
within $C'$ and the polynomial boundedness of $I_{\phi}$; 
in the second region the polynomial boundedness of 
$(1+|p|)^{N}|\widehat{\psi t_\lambda^{\alpha_1}}(p)|$ and 
the fast decay of $I_{\phi}$ (\ref{eq:oscint}). This proves 
the desired estimate for $\widehat{\phi t_{\lambda}^{\alpha_2}}$. 
\end{pf}
 
The microlocal scaling degree $\msd$ has similar properties 
as the scaling degree, as described in 
Lemma \ref{L:propsd}. In addition, one finds the following two properties: 

\begin{Lemma}\label{L:propmsd}
Let $t_1 ,t_2\in\E{D}^{\prime}(\E{M})$ with 
$\msd$ $\omega_1$ resp. $\omega_2$ at $\E{N}\subset\E{M}$, 
w.r.t. $\Gamma^1_0$ resp. 
$\Gamma^2_0$ and such that
$Z(N^\ast\E{N})\notin(\Gamma^{1}_0 +\Gamma^{2}_0 )$. 
Then the pointwise product $t_1 t_2$ exists in a small neighbourhood of 
$\E{N}$ and has 
the microlocal scaling degree $\omega\le \omega_1 +\omega_2$ at $\E{N}$
w.r.t. 
$\Gamma_0 = \Gamma^{1}_0 \cup\Gamma^{2}_0\cup(\Gamma^{1}_0 +\Gamma^{2}_0 )$. 
\end{Lemma}
\begin{pf}
By assumption, the wave front sets of 
$t_{1,\lambda}^\alpha$ and $t_{2,\lambda}^\alpha$, some $\alpha\in\E{Z}$, 
on a sufficiently small neighbourhood of $\E{N}$ satisfy
the condition 
$(\WF(t_{1,\lambda}^\alpha)+\WF(t_{2,\lambda}^\alpha))\cap 
Z(T^\ast(T_{\E{N}}\E{M}))=\emptyset$, 
hence their product exists there by {\bf M1}. 
Because of the sequential continuity of the products
in the H\"ormander topology {\bf M2}, the microlocal scaling degree is 
given by the sum w.r.t. the stated conic region as follows from {\bf M1} and 
does not depend on the choice of the map $\alpha$.
\end{pf}

The following nice property follows from the sequential continuity
of the restriction operator to submanifolds {\bf M3}:

\begin{Lemma}\label{L:bound}
Let $\E{N}_{1}$ be a submanifold of $\E{N}$, and 
let $t\in\E{D}^{\prime}(\E{M})$ have  
the microlocal scaling degree $\omega$  at $\E{N}$ w.r.t. $\Gamma_0$. 
Then the microlocal scaling degree of $t$ at $\E{N}_1$ w.r.t. the
restriction $\Gamma_1$ of $\Gamma_0$ to $\E{N}_1$ is less or equal to
$\omega$, 
$\msd^{\Gamma_1}_{\E{N}_{1}}(t)\le\msd^{\Gamma_0}_{\E{N}}(t)$. 
\end{Lemma} 

A last word is devoted to a pointlike trivialization of the above procedure.
This case can be derived straightforwardly by considering
$\E{N}\equiv \{p\}$, where $p\in\E{M}$ is a generic point and thought
of as (a rather singular case of) a submanifold.

The translation to this simpler case is done via the following 
correspondence between geometrical and analytical quantities: 

\begin{eqnarray*}
& U\in T_{\E{N}}\E{M} &\longrightarrow\qquad U_p\in T_{p}\E{M}\ ,\\
&&\null\\
& \alpha :U\rightarrow \E{N}\times\E{M} &\longrightarrow\qquad 
\alpha :U_p\rightarrow \E{M}\ ,\quad\text{$(i)$ and $(iv)$ valid},\\
&&\null\\
&t^\alpha = ({\mathbf 1}\otimes t)\circ \alpha &\longrightarrow
\qquad t^\alpha = t\circ\alpha\ ,\\
&&\null\\
&\Gamma\subset \udot{T}^{\ast} U\ ,\quad \Gamma\perp T(T\E{N}\cap U)
&\longrightarrow\qquad \Gamma\subset \udot{T}^{\ast}U_p\ ,\quad 
\Gamma\perp TU_p \ ,\\
&&\null\\
&\Gamma_0\subset \udot{N}^{\ast}\E{N}
&\longrightarrow\qquad \Gamma_p\subset \udot{T}^{\ast}_p \E{M}\ .
\end{eqnarray*}

\subsection{Transversal scaling degree}
Instead of blowing up distributions on $\E{M}$ to 
distributions on $\E{N}\times\E{M}$ in the definition of 
the scaling degree, one could also use a fixed fibration of 
a neighbourhood of $\E{N}$ in $\E{M}$ by transversal surfaces. For this 
purpose we decompose $T_{\E{N}}\E{M}$ into 
complementary subbundles, $T_{\E{N}}\E{M}=T\E{N}+C$. 
The map 
$\alpha_{C}:=\pi_{2}\circ\alpha\!\!\restriction_{C\cap V}$ , 
with the projection 
$\pi_{2}:\E{N}\times\E{M}\rightarrow\E{M}$ onto 
the second factor and with $V$ being a sufficiently small 
neighbourhood of the zero section $Z(T_{\E{N}}\E{M})$, is then a 
diffeomorphism onto some neighbourhood of 
$\E{N}$. The images of the fibers of $C$ are 
transversal surfaces. The transversally (w.r.t. $\alpha$ 
and $C$) scaled distribution is then defined by
\begin{equation}
        t_{\lambda,\bot}(x,\eta)=t\circ\alpha_{C}(x,\lambda\eta)\ ,
        \ (x,\eta)\in C \ .
        \label{E:transversal scaling}
\end{equation}
The {\it transversal} microlocal scaling degree 
may then be defined as the 
infimum of all $\omega\in\Bbb{R}$ such that the sequence
$\lambda^{\omega}t_{\lambda,\bot}$ converges to zero within 
$\E{D}^{\prime}_{\Gamma_{C}}(C)$, with a closed conical set 
$\Gamma_{C}\subset \udot{T}^{\ast}C$ with 
$\Gamma_{C}\!\!\restriction_{Z(C)}=\alpha_{C}^{\ast}(\Gamma_{0})$. 
Fortunately, it turns out that this new concept of a scaling 
degree at a surface coincides with the old one. Thus, in 
particular, the transversal scaling degree does not depend 
on the choice of the fibration.

\begin{Prop}
    The transversal microlocal scaling degree defined above 
    coincides with the microlocal scaling degree defined 
    in (\ref{E:defmsd}).
\end{Prop}
\begin{pf}
    We may restrict ourselves to a sufficiently small 
    neighbourhood of a point at the surface $\E{N}$. In 
    suitable coordinates, $\E{M}$ is a  subset of 
    $\Bbb{R}^{\delta}\times\Bbb{R}^{d-\delta}$ where the 
    first factor corresponds to $\E{N}$ and the second 
    factor to the transversal surfaces. $T_{\E{N}}\E{M}$ is 
    a subset of 
    $\Bbb{R}^{\delta}\times\Bbb{R}^{\delta}\times\Bbb{R}^{d-\delta}$
    with the first factor corresponding to $\E{N}$, the 
    second to the tangent spaces of $\E{N}$ and the third 
    one to the fibers of the transversal bundle $C$. The map $\alpha$ may 
    be chosen as
    \begin{equation}
        \alpha(x,\xi,\eta)=(x,x+\xi,\eta)
        \label{E:alpha in coordinates}\ .
    \end{equation}
    Then $\alpha_{C}$ becomes the identity. The distribution 
    $t$ may be replaced by a distribution with compact 
    support, and the factor $\bf 1$ in the blow up of $t$ 
    may be replaced by a test function $\chi\in\E{D}(\E{N})$. 
    For the Fourier transforms we then obtain
    \begin{equation}
        \widehat{t_{\lambda}}(p,q,k)=
        \lambda^{-d}
        \widehat{\chi}(p-\lambda^{-1}q)
        \widehat{t}(\lambda^{-1}q,\lambda^{-1}k)\ ,
    \end{equation}
    and 
    \begin{equation}
        \widehat{t_{\lambda,\bot}}(p,k)=
        \lambda^{d-\delta}
        \widehat{t}(p,\lambda^{-1}k)\ .
    \end{equation} 
    Furthermore, using a corresponding trivialization of 
    the respective cotangent bundles, we may identify 
    $\Gamma_{0}=\Gamma_{C}$ with 
    $\{0\}\times K$, where $K$ is a closed 
    cone in $\Bbb{R}^{d-\delta}$, considered as the 
    transversal part of the
    cotangent space, and $\Gamma$ with 
    $\{0\}\times\{0\}\times K$. 
    The convergence of $t_{\lambda}$ may be discussed in 
    terms of seminorms
    of the form $\int_{V}(1+|p|)^{N}|t_{\lambda}|$ with,
    respectively,  
    conical neighbourhoods $V$ of $\Gamma$ and some
    $N\in(-\Bbb{N})$ 
    and closed conical sets $V$ in the complement of 
    $\Gamma$ and all $N\in\Bbb{N}$. Since $\chi$ is strongly 
    decreasing, these seminorms of $t_{\lambda}$ can be estimated in terms 
    of the corresponding seminorms of $t_{\lambda,\bot}$.  
\end{pf}    

\subsection{Extension of distributions to  surfaces}

We now want to apply these concepts to the 
extension problem of distributions 
$t\in\E{D}^{\prime}(\meno{\E{M}}{\E{N}})$. The wavefront set 
of the extension shall be orthogonal to $T\E{N}$, hence a 
necessary condition is that this holds true for 
the closure of $\WF(t)$ within 
$T^{\ast}\E{M}$.  We 
extend the notion of the microlocal scaling degree to such 
distributions in an analogous way as in the extension 
problem to a single point.
Namely, for an arbitrary function 
$\chi\in\E{E}(\E{M})$ 
with $\supp\chi\cap\E{N}=\emptyset$,  
$(1\otimes\chi)\circ\alpha\cdot t^{\alpha}_{\lambda}$ 
can be considered  as a 
distribution on $U\subset T_{\E{N}}\E{M}$. 
The microlocal scaling degree of $t$ 
at $\E{N}$ is then defined in terms of all sequences 
so obtained. 

We  choose a fibration of a neighbourhood 
of $\E{N}$ by transversal surfaces.
It is easy to see that if $t$ has a  scaling 
degree $\omega$ at $\E{N}$, than its 
restriction to a transversal surface has a 
scaling degree at the point of intersection with $\E{N}$
which  is less or equal to 
$\omega$. We therefore obtain the 
corresponding extension theorem.  

\begin{Thm}\label{T:extension}
Let $\E{N}$ be a submanifold of the manifold $\E{M}$,
and let $t_0\in \E{D}^{\prime}(\meno{\E{M}}{\E{N}})$.
\begin{itemize}
\item[$(i)$] If $\sd_{\E{N}}(t_0)<
\codim(\E{N})$ 
   there exists a unique distribution $t\in\E{D}^{\prime}(\E{M})$
extending $t_0$ with $\sd_{\E{N}}(t)=\sd_{\E{N}}(t_{0})$
\item[$(ii)$] If
  $\codim(\E{N})\le\sd_{\E{N}}(t_0)<\infty$
  there exist extensions $t\in\E{D}^{\prime}(\E{M})$
  with $\sd_{\E{N}}(t)=\sd_{\E{N}}(t_0)$. They
  are uniquely characterized by their values on some closed
  subspace of $\E{D}(\E{M})$ which is complementary to the space of all
  test functions which vanish on $\E{N}$ up to order
  $\sd_{\E{N}}(t_0)-\codim(\E{N})$.  
\end{itemize}
\end{Thm}
\begin{pf}
According to Theorems \ref{T:noren} and \ref{T:reno} there exist extensions 
of the restrictions of $t_0$ to every transversal surface with the same
scaling degree. We have to show that they are restrictions of a unique
distribution $t$ on $\E{M}$ with the same microlocal scaling degree.
We first fix some normal fibration as described above 
and consider $t_{0}$ as a distribution on 
$\meno{U}{\{0\}}$ with a 
neighbourhood $U$ of the zero section of $C$. We  perform 
the construction of $t$ at all fibers, by choosing 
a smooth function $\vartheta\in\E{E}(C)$ which is 
equal to 1 in a neighbourhood of the zero section
and whose restrictions to every fiber
have compact support. Moreover, we choose smooth 
functions $\fra{w}_{\beta}\in\E{E}(U)$, also 
with compact support on each fiber, which satisfy 
the condition 
$\partial^{\gamma}_{\xi}\fra{w}_{\beta}(x,0)=\delta_{\beta}^{\gamma} $ 
where $\xi$ denotes  the variable in the fiber over 
$x\in\E{N}$. We may take 
$\fra{w}_{\beta}=\fra{w}\xi^{\beta}/\beta!$ 
with some function $\fra{w}$ which is identical 
to 1 in a neighbourhood of the zero section.

We set 
$\rho=\sd_{\E{N}}(t_0)-\codim(\E{N})$,
$\vartheta_{\lambda^{-1}}(x,\xi)=\vartheta(x,\lambda^{-1}\xi)$ 
and 
$W\phi(x,\xi)=\phi(x,\xi)-
\sum_{|\beta|\le\rho}\fra{w}_{\beta}(x,\xi)\partial^{\beta}_{\xi}\phi(x,0)$.  
Let $\Gamma_{C}=\overline{\WF(t_{0})}\cup N^{\ast}\E{N}$. 
We already 
know that the sequence 
$t_{n}=t_{0}(1-\vartheta_{2^n})\circ W$ converges on 
every fiber, and it is easy to see that it converges 
weakly in $\E{D}^{\prime}(U)$. We now want to show 
that the wave front set of $t$ is perpendicular 
to $T\E{N}$. For this purpose we show that the 
above sequence converges even in 
$\E{D}^{\prime}_{\Gamma_{C}}(U)$, i.e. that for 
every pseudodifferential operator $A$ whose 
wave front set does not intersect $\Gamma_{C}$ (hence 
$At_{n}$ is smooth) the 
sequence $At_{n}$ converges in the sense of 
smooth functions. For a pseudodifferential 
operator with smooth kernel the argument is 
essentially the same as for the weak convergence, 
hence we may restrict ourselves to 
pseudodifferential operators whose kernels have 
support in  a sufficiently small neighbourhood of 
the diagonal of $U\times U$ (only here 
singularities may occur). 

According to the discussion of condition $\bf M2$(b), we may equivalently 
look at the Fourier transform of 
$\chi t_0(\vartheta_{2^m}-\vartheta_{2^{m+1}})$ where $\chi$ 
is a test function with sufficiently small support 
which does not vanish at some point 
$x_0\in \E{N}$. Introducing suitable local coordinates in a neighbourhood of 
$x_0$, we find
\begin{equation*}
\left[ \chi t_0 ( \vartheta_{2^m} -\vartheta_{2^{m+1}} ) \right]^{\wedge}(p,k)=
2^{-m(d-\delta)}
\left[(\chi t_0 )_{2^{-m},\bot} ( \vartheta-\vartheta_2 ) 
\right]^{\wedge}(p,2^{-m}k)\ ,
\end{equation*}
where the symbol $\left[\ \cdot\ \right]^{\wedge}$ means Fourier transform, 
the $p$'s are the dual
coordinates w.r.t. points  $x\in\E{N}$ and similarly 
the $k$'s are dual coordinates w.r.t. the points 
$\xi\in C_{x}$ of the 
fibers $C_{x}$ of $C$ and finally $\delta$ is the dimension of $\E{N}$. 

By the assumption on the scaling degree of $t_0$ at $\E{N}$ we know that for 
all $\varepsilon >0$, $N\in\Bbb{N}$ and $\omega>\sd_{\E{N}}(t_0)$ there 
exists some $c>0$ such that
\begin{equation}
|[(\chi t_0)_{2^{-m},\bot}(\vartheta-\vartheta_2)]^{\wedge}(p,k)|
\le c\ 2^{m\omega}(1+|p|+|k|)^N \ ,
\label{E:locFT}
\end{equation} 
for all $(p,k)$ with $|p|>\varepsilon |k|$ (i.e. outside of a certain conical 
neighbourhood of the normal bundle $\{(p,k),p=0\}$). Therefore, the sequence
\begin{equation}
\sup_{|p|>\varepsilon|k|}
|[\chi t_0(\vartheta_{2^m}-\vartheta_{2^{m+1}})]^{\wedge}(p,k)|(1+|p|+|k|)^N
\ ,
\end{equation} 
is summable for $\omega < (d-\delta)$. If $\sd_{\E{N}}(t_0)<(d-\delta)$ such an 
$\omega$ exists, and we conclude, that the sequence $t_0(1-\vartheta_{2^m})$ 
converges in $\E{D}^{\prime}_{\Gamma_{C}}(M)$.

In case of $\sd_{\E{N}}(t_0)\ge (d -\delta)$ we have to apply the $W$-operation 
defined above, which, for $\chi$ with sufficiently small support, 
reduces to a subtraction of the 
Taylor series up to order $[\rho]$. We obtain from the integral formula 
for the remainder in the Taylor expansion 
\begin{equation*}
\begin{split}
[\chi & t_0(\vartheta_{2^m}-\vartheta_{2^{m+1}})\circ W]^{\wedge}(p,k)=\\
&\int^{1}_{0} \sum_{|\beta|=[\rho]+1} 
[\chi t_0(\vartheta_{2^m}-\vartheta_{2^{m+1}})\xi^{\beta}]^{\wedge}(p,\mu k) 
k^{\beta} \frac{(1-\mu)^{[\rho]}([\rho]+1)}{\beta!} d\mu\ .
\end{split}
\end{equation*} 
Since $\chi t_0\xi^{\beta}$ has scaling degree 
$\sd_{\E{N}}(t_0)-[\rho]-1<(d-\delta)$, we may use the same estimate as before 
and find that the sequence $t_0(1-\vartheta_{2^m})\circ W$ converges in 
$\E{D}^{\prime}_{\Gamma_{C}}(M)$.

It remains to prove the stability of the scaling degree under the extension 
procedure. The argument is a straightforward combination of the techniques in 
the corresponding proofs in Section 5 and the arguments above and is 
therefore omitted.  
\end{pf}

\section{Extension to the Diagonal and Renormalization}

We come to the main point, namely to prove 
that the inductive analysis of Section 4 closes when supplemented by the
information about the scaling degrees and gives
well defined operator-valued distributions $T_n$ all over the space $M^n$.
This process is what it is usually called ``renormalization.''

Toward this goal we must show that the scaling degree for the
distributions of the $n$th order of the induction 
can be estimated in terms of those of lower orders. 
As explained before it is sufficient to do it for 
the numerical distributions
\begin{equation}
         {^0}t(x_{1},\dots,x_{n})=\omega({^0}T(x_{1},\dots,x_{n}))\ ,
\end{equation}
where $\omega$ is our reference Hadamard state. 
According to (\ref{E:tensorp})  ${^0}t$ is a 
finite sum of terms
\begin{equation}
        f_{I}(x)t^{I}(x_{I})t^{I^c}(x_{I^c})\prod_{(i,j)\in{I\times I^c}}
        \omega_{2}(x_{i},x_{j})^{a_{ij}}\ ,
        \label{E:r}
\end{equation}
with nonnegative integers $a_{ij}$ and a smooth 
partition of unity of $\meno{M^n}{\Delta_{n}}$ 
$f_{I}\in\E{E}(\meno{M^n}{\Delta_{n}})$ with $\supp 
f_{I}\subset \C{C}_{I}$. 
$M^n$ inherits a natural metric from $M$, and all partial diagonals
$\Delta_I$ are totally geodesic submanifolds. The map
$\alpha:T{M^n}\rightarrow M^n\times M^n$ may therefore be defined by
$\alpha(x,\xi)=(x,\exp_x\xi)$. Then all restrictions of $\alpha$ to
partial subdiagonals, $\alpha_I=\alpha\!\!\restriction_{T_{\Delta_I}M^n}$
satify the conditions before Proposition 6.4.  
We can choose the functions $f_{I}$ 
with smooth scaling degree 0 at the small diagonal.
We then may consider all factors in (\ref{E:r}) as 
distributions on $M^n$. According to Lemma \ref{L:bound}, 
their microlocal scaling degrees with 
respect to $\Delta_{n}$ are bounded from above by 
their microlocal scaling degrees with respect to the 
respective partial diagonals. Moreover the convex 
combinations of the respective conical subsets of the 
conormal bundle of the small diagonal do not meet the zero 
section. Hence, by Lemma \ref{L:propmsd}, the scaling 
degree of the distribution in (\ref{E:r}) is 
bounded by
\begin{equation}
        \omega=
        \sd_{\Delta_{I}}(t_{I})+
        \sd_{\Delta_{I^c}}(t_{I^c})+
        \sum_{ij}a_{ij}(d-2)\ .
        \label{E:s}
\end{equation}
From this we get the formula
\begin{equation}
        \sd_{\Delta_{n}}
        (\omega(T(\prodwick{i=1}{n}{l_{i}}{x_{i}})))\le
        \sum_{i=1}^{n} l_{i}\frac{d-2}{2} \ .
        \label{powerc}
\end{equation}

We thus obtain
\begin{Thm}[Main Theorem] All polynomially interacting quantum field 
theories based on the the scalar field on $d\ge 2$ dimensional 
 globally hyperbolic space-times follow the same short-distance perturbative 
classification as on their respective 
Minkowskian cases.
\end{Thm}

We recall that for simplicity we have considered only
pure monomials as interacting terms, i.e. without derivatives, multiple
interacting fields and so forth. The general case however can be derived 
straightforwardly from our construction and we leave the task to the
reader.
 
We close this section by specifying the choices which have to be made
in the abstract geometrical setting of Section 6. First we choose the
normal fibration of $M^n$ in a neighbourhood of $\Delta_n$ which was
used in the proof of the Theorem 6.9 for the construction of the
extension. Let $N\Delta_n$ be the orthogonal complement of $T\Delta_n$
in $T_{\Delta_n}M^n$ w.r.t. the metric in $M^n$,
\begin{equation}
N\Delta_n=\{(x,\xi_1,\ldots,\xi_n)\in T_{\Delta_n}M^n,\sum\xi_i=0\}\ .
\end{equation}
Then $\pi_2\circ\alpha\!\!\restriction_{N\Delta_n}$ describes the desired
fibration,
\begin{equation}
\pi_2\circ\alpha(x,\xi_1,\ldots,\xi_n)=(\exp_x\xi_1,\ldots,\exp_x\xi_n)\ .
\end{equation}
$x$ may be considered as the center of mass of the points
$x_i=\exp_x\xi_i$, and the tangent vectors $\xi_i$ with the constraint
$\sum\xi_i=0$ play the r\^{o}le of relative coordinates. We further choose
a smooth function $\fra{w}$ on $TM$ which is equal to 1 on a
neighbourhood of the zero section and has compact support on each
fiber and use the function
\begin{equation}
\fra{w}_n(x,\xi_1,\ldots,\xi_n)=\prod \fra{w}(x,\xi_i)\ ,
\end{equation}
in the extension to the diagonal $\Delta_n$.

By these conventions we get a reference definition of time-ordered
products for all Wick products which involves only the chosen
Hadamard state and the function $\fra{w}$. The algebra of interacting
fields can then be defined by choosing the coefficients in the
Lagrangian. 

\section{On the Definition of the Net of Local Algebras of Observables}
In the preceding Section we finished the construction of time-ordered 
products of Wick polynomials of the free fields. We now want to show that 
this already gives the full net of local algebras of observables (within 
perturbation theory). An ``adiabatic limit,'' whatever this might mean on a 
curved space-time, is not required. Actually, these observations are not 
completely new. In \cite{B:slavnov} it was already observed that the local 
$S$-matrices of the St\"uckelberg-Bogoliubov-Epstein-Glaser approach give a 
local net of observables.

Let $\E{W}$ be the set of Wick polynomials with coefficients from $\E{D}(M)$. 
So every $A\in\E{W}$ is an operator valued distribution with compact support 
which is relatively local to the free field. Our starting relation is the 
causal factorization (\ref{b})
\begin{equation}
S(A+B+C)=S(A+B)S(B)^{-1}S(B+C)\ ,
\label{E:factorization}
\end{equation} 
for $A,B,C\in\E{W}$, whenever $\supp (A)$ is later than $\supp (C)$.

Now let $\E{L}$ be our interaction Lagrangian. Then $g\E{L}\in\E{W}$ for 
$g\in\E{D}(M)$. We may define observables with respect to the interaction 
$g\E{L}$ by Bogoliubov formula 
\begin{equation}
S_{g\E{L}}(A)=S(g\E{L})^{-1}S(g\E{L}+A)\ .
\label{E:intobs}
\end{equation}
We now show that the interacting observables depend only locally on the 
interaction. More precisely, we have the following   

\begin{Prop}
Let $\E{O}$ be a causally closed region, and let the test functions $g$ and 
$g'$ coincide on some neighbourhood of $\E{O}$. Then there exists a unitary 
$V$ such that for all $A\in\E{W}$ with $\supp (A) \subset \E{O}$ 
\begin{equation}
VS_{g\E{L}}(A)V^{-1}=S_{g'\E{L}}(A)\ .
\label{E:independence}
\end{equation}  
\end{Prop}

\begin{pf}
We may split $g'-g=a+b$ where $a$ does not intersect the past of $\E{O}$ and 
$b$ not the future. Then $\supp (a\E{L})$ is later than $\supp (A)$, hence from 
(\ref{E:factorization}) we find
\begin{equation}
S(g'\E{L}+A)=S(g'\E{L})S((g+b)\E{L})^{-1}S((g+b)\E{L}+A)\ ,
\label{E:indepfuture}
\end{equation}   
thus $S_{g'\E{L}}(A)=S_{(g+b)\E{L}}(A)$. Moreover, $\supp (A)$ is later than 
$\supp (b\E{L})$, hence 
\begin{equation}
S((g+b)\E{L}+A)=S(g\E{L}+A)S(g\E{L})^{-1}S((g+b)\E{L})\ .
\label{E:indeppast}
\end{equation}
Hence we obtain (\ref{E:independence}) with $V=S_{g\E{L}}(b\E{L})^{-1}$.  
\end{pf}

We conclude that the algebra $\fra{A}_{g\E{L}}(\E{O})$ which is generated by 
$S_{g\E{L}}(A)$, $\supp (A)\subset \E{O}$ with $g\equiv 1 $ in a neighbourhood 
of $\E{O}$, is up to unitary equivalence uniquely fixed by $\E{L}$.

We may formalize the construction in the following way. Let $\Theta(\E{O})$ 
for a causally closed compact region $\E{O}$ be the set of test 
functions which equal unity on a neighbourhood of $\E{O}$. Consider for 
$g,g'\in\Theta(\E{O})$ the set $\E{V}_{gg'}(\E{O})$ of unitaries $V$ which 
satisfy the intertwining relation
\begin{equation}
VS_{g\E{L}}(A)=S_{g'\E{L}}(A)V \ ,\ \supp (A)\subset\E{O} \ .
\label{E:intertwining}
\end{equation}
The algebra of observables $\fra{A}_{\E{L}}(\E{O})$ can now be defined as the 
algebra of covariantly constant sections of the bundle
\begin{equation}
\bigcup_{g\in\Theta(\E{O})}\{g\}\times\fra{A}_{g\E{L}}(\E{O}) \ .
\label{E:obsbundle}
\end{equation}
Here, a section $A=(A_g)_{g\in\Theta(\E{O})}$ is called covariantly 
constant if
\begin{equation}
VA_g=A_{g'}V\ , \qquad\forall V\in \E{V}_{gg'}(\E{O}) \ .
\label{E:covconst}
\end{equation} 
$\fra{A}_{\E{L}}(\E{O})$ contains for example the elements 
$S_{\E{L}}(A)=(S_{g\E{L}}(A))_{g\in\Theta(\E{O})}$. 
To complete the construction of the net of local algebras of observables we 
have to fix the imbeddings 
$i_{\E{O}_2\E{O}_1}:\fra{A}_{\E{L}}(\E{O}_1) \rightarrow 
\fra{A}_{\E{L}}(\E{O}_2)$ for $\E{O}_1\subset\E{O}_2$. 
But this structure is inherited from the fibers 
and may be defined by the restriction of the section from $\Theta(\E{O}_1)$ 
to $\Theta(\E{O}_2)$.     

\section{Summary and Outlook}
We have proven that renormalization on curved backgrounds can be done in close
analogy to renormalization on Minkowski space, and that the removal
of singularities follows the well known power counting rules.
This result was expected since the ultraviolet behaviour on
smooth manifolds should be essentially identical to that
on Minkowski space. But we had to overcome two major
obstacles: On a generic space-time there is no reason to
expect a decent infrared behaviour, hence we had to use a method
which decouples completely the short distance from the
long distance problem; the other source of problem
was the absence of translation invariance which, on the technical side
makes obsolete the usual momentum space methods, and on the side of physics, 
forbids to base the construction on a distinguished vacuum state and on the
notion of particles.

We solved these problems by basing the construction on the local $S$-matrices
of St\"uckelberg and Bogoliubov, by invoking the general ideas of 
algebraic quantum field theory and by replacing translation invariance
by smoothness properties, making extensive use of techniques and concepts
from microlocal analysis. 

Besides the solution of the problem to which this paper is addressed, we 
solved several other problems which might be of independent interest. First
we have found a new construction of Wick polynomials on a 
domain which depends only
on the representation associated to a fixed Hadamard state but not on the state
itself. Since according to Verch \cite{B:verch}, all 
representations induced by Hadamard states 
are locally quasiequivalent this amounts to an algebraic 
construction of Wick polynomials. 

Second, we gave a perturbative construction 
of algebraic quantum field theory.
In particular, we proved that the theory (in the algebraic sense), is 
completely fixed if it is known locally. Actually this holds independently 
of perturbation theory and might be a hint for a construction 
(in the sense of constructive quantum field theory) 
of asymptotically free theories. There the construction in small volumes
seems to be possible \cite{B:mrs} but the infrared problem poses, at present,
unsormountable difficulties. The message of this paper is that the construction
of the algebra of observables is nevertheless possible once the algebra of
observables for small space-time regions have been constructed. The long 
distance behaviour of such a theory would still require an extra 
investigation, but it would be the behaviour of an existing theory, 
quite similar to the computation of spectra of Hamiltonians which have been
shown to be self-adjoint operators.

On the technical side we had to study the extension problem for distributions
which are defined on the complement of some submanifold. This seems to be a 
natural mathematical question, and in a simple case it was treated in
\cite{B:estrada} by similar methods. What seems to be new is our concept
of the (microlocal) scaling degree at a surface which combines the condition 
of smoothness along a surface with a classification of the singularity in a
transversal direction.

The main open point in this paper is the fixing of the finite 
renormalizations.
One expects that they can be chosen in terms of local functions of the metric,
but a precise formulation meets a lot of problems. A similar problem was
studied (and partially solved) in the definition of the expectation value
of a renormalized energy momentum tensor of free fields by R.~Wald 
\cite{B:wald2}.
We hope to return to this problem in a future publication \cite{B:bf2}.

\section*{Acknowledgements}
We are particularly grateful to Raymond Stora who long ago suggested us
the relevance of the Epstein and Glaser procedure for 
renormalization on curved backgrounds. 
In an early stage of this work, some results, in particular 
on the wave front sets of time-ordered functions, have been 
obtained in collaboration with Martin K\"ohler which is 
gratefully acknowledged.
The first named author was partially supported by a grant of Training 
and Mobility of Researchers (TMR) programme of European Community.


\begin{thebibliography}{999}

\bibitem{B:ashtekar} Ashtekar,~A.: Mathematical problems of non-perturbative
quantum general relativity. In: Zinn-Justin et al. (eds.) 
Les Houches Summer School on Gravitation and Quantization.  
North-Holland, 1994

\bibitem{B:bee} Beem,~J.~K., Ehlrich,~P.~E. and Easley,~K.~L.: 
Global Lorentzian Geometry. New York: Marcel Dekker Inc., 1996

\bibitem{B:birrel} Birrel,~N.~D., and Davies,~P.~C.~W.: Quantum 
Fields in Curved Space. Cambridge: Cambridge University Press, 1982

\bibitem{B:bs} Blanchard,~P., and S\'en\'eor,~R.: Green's functions for 
theories with massless particles (in perturbation theory). Ann. Inst. Henry 
Poincar\'e {\bf 23}, 147 (1975)

\bibitem{B:bos} 
Bogoliubov,~N.~N., and Shirkov,~D.~V.: Introduction to the 
Theory of Quantized Fiels. New York: John Wiley and Sons, 1976, 3rd edition

\bibitem{B:bourbaki} Bourbaki,~N.: Alg\`ebre, Chap.VIII. Paris: Hermann, 1970

\bibitem{B:bem} Bros,~J., Epstein,~H., and Moschella,~U.: Analyticity 
properties and thermal effects for general quantum field theory on de Sitter
space-time. Commun. Math. Phys. {\bf 186}, 535 (1998)

\bibitem{B:bfk} Brunetti,~R., Fredenhagen,~K., and K\"ohler,~M.:
The microlocal spectrum condition and the Wick's polynomials 
of free fields. Commun. Math. Phys. {\bf 180}, 633 (1996)

\bibitem{B:bfroma} Brunetti,~R., and Fredenhagen,~K.: 
Microlocal analysis and interacting quantum field theory: 
Renormalizability of $\varphi^4$. In: Doplicher,~S., Longo,~R., 
Roberts,~J.~E., and Zsido,~L (eds.)
Operator Algebras and Quantum Field Theory. Proceedings, Roma 1996, 
International Press 1997

\bibitem{B:bf2} Brunetti,~R., and Fredenhagen,~K., work in progress

\bibitem{B:bfrev} Brunetti,~R., and Fredenhagen,~K.: On the connection
between interacting quantum field theory and microlocal analysis. 
Forthcoming review paper

\bibitem{B:buchholz} Buchholz,~D.: Current trends in axiomatic 
quantum field theory. {\tt hep-th/9811233}

\bibitem{B:bunch} Bunch,~T.~S.: BPHZ Renormalization of $\lambda\phi^4$ 
field theory in curved spac-times. Ann. of Phys. {\bf 131}, 118 (1981)

\bibitem{B:dwb} De Witt,~B.~S., and Brehme,~R.~W.: Radiation damping in 
a gravitational field. Ann. of Phys. {\bf 9}, 220 (1965)

\bibitem{B:dimock} Dimock,~J.: Scalar quantum field in an external 
gravitational field. J. Math. Phys. {\bf 20}, 2549 (1979)

\bibitem{B:dm} Dosch,~H.~G., and M\"uller,~V.~F.:
Renormalization of quantum electrodynamics in an arbitrary strong time 
independent external field. Fort. der Physik {\bf 23}, 661 (1975)

\bibitem{B:dh} Duistermaat,~J.~J., and H\"ormander, L.: 
Fourier integral operators II. Acta Math. {\bf 128}, 183 (1973)

\bibitem{B:df} D\"utsch,~M., and Fredenhagen,~K.: A local (perturbative)
construction of observables in gauge theories: the example of QED.
 To appear on Commun. Math. Phys. (1999)

\bibitem{B:dyson} Dyson,~F.: Collected works. American Mathematical 
Society. Providence RI: International Press, 1996

\bibitem{B:eg} Epstein,~H., and Glaser,~V.: The role of locality in 
perturbation theory. Ann. Inst. Henri Poincar\'e-Section A, vol. XIX, 
n.3, 211 (1973)

\bibitem{B:eginfrared} Epstein,~H., and Glaser,~V.: Adiabatic limit in 
perturbation theory. In: Velo,~G., and 
Wightman,~A.~S. (eds.) Renormalization Theory. Proceedings, 
D. Reidel Publishing Co., Dodrecht-Holland, 1976

\bibitem{B:es} Epstein,~H.: On the Borchers class of a free field.
Nuovo Cimento {\bf 27}, 886 (1966)

\bibitem{B:estrada} Estrada,~R.: Regularization of distributions.
Internat. J. Math. \& Math. Sci. {\bf 21}, 625 (1998)

\bibitem{B:fh} Fredenhagen,~K., and Haag,~R.: Generally covariant quantum 
field theory. Commun. Math. Phys. {\bf 108}, 91 (1987)

\bibitem{B:fulling} Fulling,~S.: Aspects of Quantum 
Field Theory in Curved Space-Time. Cambridge: Cambridge 
University Press, 1989

\bibitem{B:gj} Glimm,~J., and Jaffe,~A.: Quantum Physics: A Functional 
Integral Point of View. New York, Berlin, Heidelberg: Springer-Verlag, 1981

\bibitem{B:gp} Guillemin,~V., and Pollack,~A.: Differential Topology.
Englewood-Cliffs, N.J.: Prentice-Hall, Inc., 1974

\bibitem{B:gsw} Green,~D.~B., Schwarz,~J.~H., and Witten,~E.:  
Superstring Theory. Voll. 1 and 2. Cambridge: 
Cambridge University Press, 1987

\bibitem{B:haag} Haag,~R.: Local Quantum Physics: Fields, particles 
and algebras. Berlin: Springer-Verlag, 2nd ed., 1996

\bibitem{B:hns} Haag,~R., Narnhofer,~H., and Stein,~U.: On quantum field 
theory in gravitational background. Commun. Math. Phys. {\bf 94}, 
219 (1984)

\bibitem{B:smbook} Halzen,~F., and Martin,~A.~D.: Quarks and Leptons: An
Introductory Course in Modern Particle Physics. New York: John Wiley and Sons, 
1984

\bibitem{B:hawking} Hawking,~S.: Particle creation by black holes. 
Commun. Math. Phys. {\bf 43}, 199 (1975)

\bibitem{B:hawkingcpc} Hawking,~S.: The Chronology protection conjecture.
Phys. Rev. {\bf D 46}, 603 (1992)

\bibitem{B:hepp} Hepp,~K.: Th\'eorie de la Renormalisation. Lect. 
Notes in Phys. {\bf 2}. Berlin, Heidelberg: Springer-Verlag, 1969

\bibitem{B:micro} H\"ormander,~L.: The Analysis of Linear Partial 
Differential Operators. Voll. I-IV. Berlin: Springer-Verlag, 1983-1986

\bibitem{B:iagolnitzerb} Iagolnitzer,~D.: Scattering in Quantum Field 
Theories: The Axiomatic and Constructive Approaches. Princeton NJ: 
Princeton University Press, 1993

\bibitem{B:iagolnitzerp} Iagolnitzer,~D.: Microlocal analysis and phase
space decomposition. Lett. Math. Phys. {\bf 21}, 323 (1991)

\bibitem{B:iz} Itzykson,~C., and Zuber,~J.~B.: Quantum Field Theory. 
New-York: McGraw-Hill, 1980

\bibitem{B:junker} Junker,~W.: Hadamard states, adiabatic vacua and the 
construction of physical states for scalar quantum fields on curved 
spacetimes. Rev. Math. Phys. {\bf 8}, 1091 (1996)

\bibitem{B:krw} Kay,~B.~S., Radzikowski,~M., and Wald,~R.~M.: Quantum 
field theories on spacetimes with a compactly generated Cauchy horizon. 
Commun. Math. Phys. {\bf 183}, 533 (1997)

\bibitem{B:kw} Kay,~B.~S., and Wald,~R.~M.: Theorems on the uniqueness and 
thermal properties of stationary, non singular, quasifree states on 
spacetimes with a bifurcate Killing horizon. Phys. Rep. 
{\bf 207}, 49 (1991)

\bibitem{B:reviewqed} Kinoshita,~T. (ed.): Quantum Electrodynamics.
Singapore: World Scientific, 1990

\bibitem{B:koehler} K\"ohler,~M.: Ph.D. thesis, University of 
Hamburg 1994 

\bibitem{B:liess} Liess,~O.: Conical refractions and higher 
microlocalization. Lect. Notes in Math. {\bf 1555}. Berlin:
Springer-Verlag, 1993
 
\bibitem{B:luescher} L\"uscher,~M.: Dimensional regularization in the 
presence of large background fields. Ann. of Phys. {\bf 142}, 
359 (1982)

\bibitem{B:mrs} Magnen,~J., Rivasseau,~V., and S\'en\'eor,~R.: Construction
of YM-4 with an infrared cutoff. Commun. Math. Phys. {\bf 155}, 325 (1993)

\bibitem{B:os} Osterwalder,~K. and Schrader,~R.: Axioms for Euclidean 
Green's functions: I, II. Commun. Math. Phys. {\bf 31}, 81 (1973);
ibidem {\bf 42}, 281 (1975)

\bibitem{B:prange} Prange, ~D.: Causal perturbation theory and differential 
renormalization. {\tt hep-th/9710225}

\bibitem{B:radman} Radzikowski,~M.: Micro-local approach to the Hadamard 
condition in quantum field theory on curved space-time. Commun. 
Math. Phys. {\bf 179}, 529 (1996)

\bibitem{B:scharf} Scharf,~G.: Finite Quantum Electrodynamics: The Causal 
Approach. Berlin: Springer-Verlag, 1995, 2nd edition

\bibitem{B:schwingerbook} Schwinger,~J. (ed.): Selected Papers on 
Quantum Electrodynamics. New York: Dover, 1960

\bibitem{B:slavnov} Il'in,~V.~A., 
and Slavnov,~D.~A.: Observable algebras in the $S$-matrix approach. 
Theor. Math. Phys. {\bf 36}, 32 (1978)

\bibitem{B:steinmann} Steinmann,~O.: Perturbation Expansions in 
Axiomatic Field Theory. Lect. Notes in Phys. 
{\bf 11}. Berlin: Springer-Verlag, 1971

\bibitem{B:stora} Stora,~R.: Differential algebras in 
Lagrangean field theory.
ETH Lectures, January-February 1993. Manuscript

\bibitem{B:sw} Streater,~R.~F., and Wightman,~A.~S.: PCT, Spin $\And$ 
Statistics and all that. New York: W.A. Benjamin, Inc., 1964

\bibitem{B:stuck} St\"uckelberg,~E.~C.~G., and Peterman,~A.: La 
normalisation des constants dans la theorie des quanta. Helv. Phys. Acta 
{\bf 26}, 499 (1953); and earlier references therein

\bibitem{B:verch} Verch,~R.: Local definitenss, primarity and 
qua\-si\-equi\-va\-len\-ce of 
qua\-si\-free Ha\-da\-mard quantum states in curved spacetime. 
Commun. Math. Phys. {\bf 160}, 507 (1994)

\bibitem{B:verch2} Verch,~R.: Wavefront sets in algebraic 
quantum field theory. {\tt math-ph/9807022}

\bibitem{B:waldgr} Wald,~R.~M.: General Relaitvity. Chicago:
The University of Chicago Press, 1984

\bibitem{B:wald} Wald,~R.~M.: Quantum Field Theory in Curved Spacetime 
and Black Hole Thermodynamics. Chicago: The University of Chicago 
Press, 1994

\bibitem{B:wald2} Wald,~R.~M.: The back reaction effect in particle 
creation in curved spacetime. Commun. Math. Phys. {\bf 54}, 1 (1977)

\bibitem{B:weinberg} Weinberg,~S.: The Quantum Theory of Fields. 
vol.I-II. Cambridge: Cambridge University Press, 1995-1996

\bibitem{B:wrezinski} Wrezinski, ~W.~F.: Note on the construction of the
Bogolyubov scattering operator in the $(:\varphi^4 :)_2$ theory. Theor. 
Math. Phys. {\bf 11}, 331 (1972)

\bibitem{B:zimmermann} Zimmermann,~W.: Convergence of Bogoliubov 
method of renormalization in momentum space. Commun. Math. Phys. {\bf 15},
208 (1969)
 
\end{thebibliography}
\end{document}